\documentclass[11pt,a4paper]{article}
\pdfoutput=1

\usepackage{jheppub}
\usepackage{latexsym}
\usepackage{multirow}
\usepackage{color}
\usepackage[usenames,dvipsnames,table]{xcolor}
\usepackage{graphicx}
\usepackage{epsfig}  
\usepackage{epsf}    
\usepackage{dcolumn}
\usepackage{bm}
\usepackage{dcolumn}
\usepackage{textcomp}
\usepackage{float}
\usepackage{subfig}
\usepackage{hypcap}
\usepackage[]{hyperref}
\usepackage{makecell}
\usepackage{color}
\usepackage{pifont}
\usepackage{appendix}
\usepackage{url}
\usepackage{cancel}
\usepackage[normalem]{ulem}

\hypersetup{
  bookmarks=true,         
  unicode=false,          
  pdftoolbar=true,        
 pdfmenubar=true,        
 pdffitwindow=true,     
 pdfstartview={FitH},    
 pdfsubject={Neutrino Oscillations Phenomenology},   
 pdfnewwindow=true,      
 pdfcreator={RevTeX},
 colorlinks=true,       
 linkcolor=red,          
 citecolor=blue,        
 filecolor=black,      
 urlcolor=blue,           
  }

\newcommand{\be}{\begin{equation}}
\newcommand{\ee}{\end{equation}}
\newcommand{\ba}{\begin{eqnarray}}
\newcommand{\ea}{\end{eqnarray}}

\makeatletter
\renewcommand{\fnum@table}{\textbf{\tablename~\thetable}}
\renewcommand{\fnum@figure}{\textbf{\figurename~\thefigure}}
\makeatother

\preprint{TIFR/TH/17-16}

\title{Constraining compressed versions of MUED and MSSM using soft tracks at the LHC} 

\author[a]{Sabyasachi Chakraborty,}
\author[b]{Saurabh Niyogi,}
\author[a]{K. Sridhar}

\affiliation[a]{Department of Theoretical Physics, Tata Institute of Fundamental Research, \\ 1, Homi Bhabha Road, Mumbai 400005, India}
\affiliation[b]{Department of Physics and Astrophysics, University of Delhi, New Delhi 110007, India}

\emailAdd{sabya@theory.tifr.res.in}
\emailAdd{saurabhphys@gmail.com}
\emailAdd{sridhar@theory.tifr.res.in}

\abstract
{A compressed spectrum is an anticipated hideout for many beyond standard model scenarios. Such a spectrum naturally arises in the minimal universal extra dimension framework and also in supersymmetric scenarios. Low $p_T$ leptons and jets are characteristic features of such situations. Hence, a monojet with $\cancel{E_T}$ has been the conventional signal at the Large Hadron Collider (LHC). However, we stress that inclusion of $p_T$-binned track observables from such soft objects provide very efficient discrimination of new physics signals against various SM backgrounds. We consider two benchmark points each for minimal universal extra dimension (MUED) and minimal supersymmetric standard model (MSSM) scenarios. We perform a detailed cut-based and multivariate analysis (MVA) to show that the new physics parameter space can be probed in the ongoing run of LHC at 13 TeV center-of-mass energy with an integrated luminosity $\sim$ 20-50 fb$^{-1}$. When studied in conjunction with the dark matter relic density constraint assuming standard cosmology, we find that compressed MUED (with $\Lambda R=2$) can be already excluded from the existing data. Also, MVA turns out to be a better technique than regular cut-based analysis since tracks provide uncorrelated observables which would extract more information from an event.}
\begin{document}
\maketitle
\flushbottom
\newpage
\section{Introduction}
In the pursuit of new physics, the two CERN-based experiments namely, ATLAS and CMS have constrained the parameter space of many beyond standard model (BSM) scenarios. However, if the spectrum of new physics particles are compressed then such stringent constraints can be somewhat circumvented. The SM jets and leptons emanating from the cascades of such compressed spectra are too soft to give rise to any reconstructed object. In such a circumstance, one has to rely on a jet recoiling against the system giving rise to a monojet + missing energy ($\cancel{E_T}$) signature at the collider assuming the lightest particle is stable within the collider. From the theoretical perspective, such a spectrum can be obtained in low-scale supersymmetric (SUSY) scenarios~\cite{Martin:2007gf,Fan:2011yu,Murayama:2012jh} and in the minimal universal  extra-dimensional framework (MUED)~\cite{Murayama:2011hj,Choudhury:2016tff}. 

On the experimental front, most of the searches performed by both ATLAS and CMS on compressed spectra are confined to SUSY scenarios. However, in all such cases, the experimental limits are rather relaxed for compressed scenarios. For example, in the framework of the minimal supersymmetric standard model (MSSM), both ATLAS and CMS have looked into the pair production of squarks ($\widetilde q_1^0$) and gluinos ($\widetilde g$) and their subsequent decays to quarks and neutralino ($\widetilde\chi_1^0$) which is also assumed to be the lightest supersymmetric particle (LSP). In such cases, the typical final state consists of 2 (4) jets with $\cancel{E_T}$. The presence of gauginos in the cascade can also increase the  jet multiplicity in an event. Therefore, the experiments carefully look into the prospect of having 2-6 jets associated with missing energy in the final state. When the $\widetilde g/\widetilde q$ and $\widetilde\chi_1^0$ masses are well separated then the bounds on squark/gluino masses are as stringent as 2 TeV~\cite{Aaboud:2016zdn,Khachatryan:2016kdk}. However, in compressed scenarios, this bound reduces to a rather relaxed value of 600~GeV~\cite{Aad:2015zva}. The situation is somewhat similar for MUED also. An important point to note in this context is that the conventional multijet+$\cancel{E_T}$ may be more useful than the prototypical monojet+$\cancel{E_T}$ searches~\cite{Dutta:2015exw}

It is only recently that the study of phenomenology of compressed supersymmetry has been taken up. In particular, coannihilation of dark matter calls for compressed spectrum at least in the dark matter sector. The phenomenology of such scenarios has been discussed in earlier works \cite{Baer:2007uz,Martin:2008aw,LeCompte:2011cn,LeCompte:2011fh,Harigaya:2014dwa,Ellis:2015vaa,Han:2014kaa,Nagata:2015hha,Bramante:2015una,Han:2014xoa,Baer:2014kya,Dutta:2017jpe,Aboubrahim:2017aen}. As LHC fails to reveal any signal of new physics, the focus tends to shift to spectra that can hide the expected signals within the experimental uncertainty. Various propositions are put forward in order to gain more sensitivity in those regions of parameter space \cite{Bhattacherjee:2013wna,Han:2013usa,Schwaller:2013baa,Barducci:2015ffa}. Jet substructure techniques and kinematic correlations between jets coming from initial state radiation also provide alternate search techniques \cite{Han:2015lha,Mukhopadhyay:2014dsa}.

However, such search strategies are rather limited as it fails to incorporate all the information from an event. Any well-motivated new physics scenario is expected to have particles ranging from the electroweak scale to the TeV scale. The presence of such particles will definitely increase the particle multiplicity in the final state of an event. If the whole spectrum is compressed then instead of giving rise to reconstructed objects, the SM partons would leave tracks in the tracker or in the muon spectrometer. A simple counting of number of tracks which is not a part of any reconstructed object thus gives a hint about the particle multiplicity in an event. Using such information, one can enhance the sensitivity of monojet+$\cancel{E_T}$ search considerably~\cite{Chakraborty:2016qim}. It is important to note that the traditional variables generally used to differentiate signal and SM background are all energy weighted. A simple counting of number of soft tracks therefore should be totally uncorrelated with the former variables. This, in principle, should work tremendously well in a multivariate analysis (MVA) as will be shown in this work. We therefore propose to make a careful comparison between the cut-based analysis and MVA and show the incorporation of the soft tracks leads to an increase in the purity and significance of the signal. Our analysis is based on 13 TeV run of LHC and hence, we expect that our strategy can work out with the already accumulated data at the LHC.

The paper is organized as follows. We start with a brief description of the two BSM scenarios, namely, MUED and MSSM in sec.~\ref{mued} and \ref{mssm}, respectively. In sec.~\ref{benchmark}, we discuss our selection of benchmark points. We present a detailed analysis (both cut-based and multivariate) and results in sec.~\ref{analysis}, and in sec.~\ref{conclu} we present our conclusion.
\section{Minimal Universal Extra Dimension \label{mued}}
In the universal extra dimension (UED) scenario, one considers an extra, flat  spatial dimension (denoted by $x_5$) which is compactified on $S^1/Z_2$ orbifold with the compactification radius $R$ \cite{Appelquist:2000nn}. The gauge structure and the particle content of the SM have been kept intact. All SM particles are allowed to propagate into bulk. The 5D gauge couplings have negative mass dimensions, hence, the theory is fundamentally non-renormalizable from dimensional argument. Therefore, the theory is expected to remain valid up to a certain scale $\Lambda$. In the minimal version of UED scenario, there are only three input parameters: the compactification radius ($R$), the cut-off scale ($\Lambda$) and the Higgs mass ($m_H$). Due to the compactification of the extra dimension, fields are periodic in $x_5$ direction and satisfy periodic boundary condition: $\Phi(x, x_5) = \Phi(x, x_5 + 2\pi R)$ and this results in the expansion of 5D fields into an infinite series of Kaluza-Klein (KK) modes. The 0-th mode is identified with the corresponding SM particle. Such a tower of infinite KK states appears for each SM particles in 4D. The KK modes carry the exactly same quantum numbers as the corresponding SM fields with some important differences. In particular, the KK fermions are vector-like states, {\it i.e.}, there are both $SU(2)$ doublet as well as singlet KK fermions in each level. We denote  doublet and singlet KK fermions by upper and lower-case symbols respectively. KK gauge bosons are denoted by upper-case symbols unless mentioned specifically.

The mass $m_n$ of $n$-th KK states is given by $m^2_n = m^2_0 + \frac{n^2}{R^2}$ where $m_0$ is the mass of the 0-th mode {\it i.e.,} the corresponding SM particle. For large $R^{-1}$,  $m^2_0 \ll \frac{n^2}{R^2}$; hence, all the KK masses at a given level $n$ are almost degenerate. 
However, masses of the KK states receive additional contributions from radiative corrections. The one-loop correction includes
both bulk and boundary contributions \cite{Cheng:2002iz}. The bulk corrections involve one-loop diagrams where the internal loop momenta
run around the compactified dimension. These corrections are finite and independent of the cut-off scale. On the other hand, one-loop boundary corrections appear at the orbifold fixed points. These corrections are logarithmically divergent which are determined from the running between the cut-off $\Lambda$ and $R^{-1}$.
It is known from ref. \cite{Cheng:2002iz} that the presence of large boundary terms would significantly affect KK masses and mixings\footnote{MUED with large boundary terms can produce a 
completely different kind of spectra and has been discussed in detail in \cite{delAguila:2003bh,Flacke:2008ne,Datta:2012xy,Datta:2012tv,Flacke:2013pla,Flacke:2014jwa}.}. While it is expected that these boundary terms should be present; however,
it is not inconsistent to assume that they are negligibly small at the scale $\Lambda$. In MUED, the boundary terms are indeed assumed to be small at $\Lambda$ and we will work with this assumption in this paper.
The radiative corrections for the KK gluon is the largest simply because of its large coupling strength and multiplicative color factor. More explicitly, we note down the non-zero bulk corrections for the gauge bosons as:
\begin{eqnarray}
\text{gluons:}~~\delta m^2(g^\mu_1) &=&  \frac{3g^2_s}{2} \frac{\zeta(3)}{16\pi^4R^2} + \frac{23g^2_s}{2} \frac{1}{32\pi^2R^2} \text{log}\left( \Lambda^2 R^2 \right), \nonumber \\
\text{W bosons:}~~\delta m^2(W^\mu_1) &=& -\frac{5}{2} \frac{g^2_w \zeta(3)}{16\pi^2R^2} + \frac{15g^2_w}{32\pi^2R^2} \text{log} \left( \Lambda^2 R^2 \right), \nonumber \\
\text{Photon:}~~\delta m^2(B^\mu_1) &=& -\frac{39}{2} \frac{g^2_1 \zeta(3)}{16\pi^2R^2} - \frac{g^2_1}{96\pi^2R^2} \text{log} \left( \Lambda^2 R^2 \right),
\end{eqnarray}
where $\zeta(3) \simeq 1.2$ and $g_s$, $g_w$, $g_1$ are the gauge coupling corresponding to the gauge groups SU(3)$_C$, SU(2)$_L$, U(1)$_Y$ respectively. The resulting mass splitting is just enough to allow cascade decays to take place. In the MUED scenario, with no boundary terms, the spectrum is completely fixed by the radius of compactification and the cut-off $\Lambda$.

One of the interesting features of the UED scenario is conservation of momentum along the extra dimensions which, in turn, leads to conservation of KK number. However, this no longer holds true once loop corrections are taken into account as the compactness of the extra dimension leads to violations of Lorentz symmetry. Further, imposing orbifold boundary conditions in order to remove unwanted degrees of freedom, breaks conservation of KK number. But KK parity (a $Z_2$
symmetry), defined as $(-1)^{n}$ where $n$ is the KK level, still remains conserved. Any odd level KK mode must be produced in association with another odd KK parity mode at the collider experiments. Conservation of KK parity also leaves the lightest KK particle (LKP) stable and if, weakly interacting, becomes a good candidate for dark matter. The LKP is the level-1 KK photon ($\gamma_1$) in the MUED scenario and generally serves as an excellent candidate for thermal weakly interacting massive particle (WIMP) \cite{Cheng:2002ej,Servant:2002aq,Hooper:2007qk}. However, severe constraints stem from relic density of the dark matter. Assuming standard cosmology, the relic density of $\gamma_1$ can be roughly approximated as,
\begin{eqnarray}
\Omega(\gamma_1) h^2 &\simeq& \frac{10^8}{M_P}\frac{x_F^2}{ax_F+3b},
\end{eqnarray}
where $x_F$ is the freeze-out epoch (usually lies in between $\sim 20-25$) and can be obtained iteratively. The self-annihilation rate of $\gamma_1$ leads to
\begin{eqnarray}
a\sim \frac{380\pi\alpha^2}{81\cos^4\theta_W^2}R^2, \hskip 1cm b\sim 0.
\end{eqnarray}
When subjected to the constraint regarding the observed value of the relic density measurement of the recent Planck data~\cite{Ade:2015xua}, it immediately puts a strong bound on the radius of compactification as~\cite{Sridhar}
\begin{eqnarray}
1.40~\text{TeV} \leqslant R^{-1} \leqslant 1.46~\text{TeV}.
\end{eqnarray}

Various collider signatures of MUED have been studied extensively over many years \cite{Bhattacharyya:2009br,Bhattacherjee:2010vm,Datta:2011vg,Belyaev:2012ai,Belanger:2012mc,Kakuda:2013kba,Dey:2014ana}. One usually looks for missing transverse energy (coming from LKP) accompanied by decay products from KK particles in the cascade. The $p_T$ of the jets/leptons is fixed by the mass splitting which, in turn, is determined by the cut-off scale $\Lambda$. A higher value of $\Lambda$ result in a larger mass splitting which ensures an easier choice of cuts and other wider search strategy to be employed. However, to prevent the scalar potential from being unbounded from below, a lower value of $\Lambda$ is preferred ($\Lambda R\lesssim 4$)~\cite{Datta:2012db}. Such a choice results in a rather difficult situation where the spectrum is squeezed. In this case, the $p_T$ of the final state jets/leptons or missing $p_T$  are, in general, small and may not pass the selection cuts. Bounds on MUED parameters from LHC run-I and II data are recently discussed in \cite{Choudhury:2016tff,Deutschmann:2017bth}. It has been concluded that monojet and multijet $+ \cancel{E_T}$ channels seem to be the best channels to probe for such a scenario \cite{Choudhury:2016tff,Khachatryan:2014rra,Aaboud:2016tnv}. In this work, we choose $\Lambda R=2$ which gives rise to very closely spaced KK states. 
\section{Minimal Supersymmetric Standard Model \label{mssm}}
In case of SUSY, the spectrum depends on the choice of SUSY breaking and mediation mechanism. For a high scale SUSY theory, renormalization group (RG) evolution equations are bound to generate a large mass splitting between the colored and non-colored superpartners. Nevertheless, any low scale mediation mechanism, like the, Scherk-Schwarz~\cite{Scherk:1978ta,Scherk:1979zr} mechanism can generate a rather compressed spectrum where the effect of RG running is minuscule. In this work, we will not bias ourselves with any particular SUSY breaking mechanism and consider all the soft masses of the sparticles to be at the low scale. However, the choice of the SUSY spectrum should be consistent with the Higgs mass as well as DM relic density and direct detection constraints. 

It is well known that in MSSM, the tree level mass of the lightest CP-even Higgs boson is bounded from above by the $Z$-boson mass. Dominant one-loop contribution mainly from the top-squarks lift the mass of the Higgs boson. In absence of any additional $F$-term~({\it{e.g.}}, NMSSM~\cite{Ellwanger:2009dp} or the next to minimal supersymmetric standard model),  or $D$-term~({\it{e.g.}}, $U(1)_X$ extended MSSM)~\cite{Batra:2003nj} contributions to the Higgs quartic term, the Higgs mass can be well approximated as~\cite{Martin:1997ns}
\begin{eqnarray}
m_h^2 &\simeq& m_Z^2 \cos^2 2\beta + \frac{3 m_t^4}{4\pi^2 v^2}\left(\ln\frac{M_S^2}{m_t^2}+\frac{X_t^2}{M_S^2}-\frac{X_t^4}{12 M_S^2}\right),
\end{eqnarray}
where $\tan\beta=v_u/v_d$, {\it i.e.}, the ratio of the up and down-type Higgs field vev's. $M_S=\sqrt{m_{\widetilde t_1} m_{\widetilde t_2}}$ and the mixing in top squark sector is parametrised by $X_t =A_t-\mu\cot\beta$. $A_t$ is the trilinear scalar coupling consisting the Higgs and top squark fields and $\mu$ is the Higgsino mass parameter. One generally requires multi-TeV top squark masses or large $A_t$ to fit the Higgs mass~\cite{Martin:1997ns}. Depending on the choices of other soft SUSY breaking parameters, we can obtain a compressed spectrum. Such a spectrum is also useful from the point of view of DM relic density as enhanced coannihilation rates would naturally increase the DM interactions thus reducing the relic density. To reiterate, such compressed scenarios are hard to probe at the LHC because of the absence of hard leptons or jets in the final state which are essential in the usual SUSY search strategies. The lack of visible energy in the final state topology relaxes the constraints on the superpartner masses considerably. As a result, compressed SUSY may turn out to be one of the last explanations for the non-observation of superpartners at the LHC.

In our analysis we choose the lightest neutralino to be the DM candidate. If it is bino-like then the annihilation cross section and its contribution to the relic density is well-approximated by~\cite{ArkaniHamed:2006mb}
\begin{eqnarray}
\langle \sigma_{\widetilde b}v\rangle &=& \frac{3 g^4 \tan^4\theta_W r (1+r^2)}{2\pi m^2_{\widetilde l_R} x (1+r)^4}, \hskip 1cm x \equiv \frac{M_1}{T}, \hskip 1cm r \equiv \frac{M_1^2}{m^2_{\widetilde l_R}}, \nonumber \\
\Omega_{\widetilde b} h^2 &\simeq& 1.3\times 10^{-2}\left(\frac{m_{\widetilde l_R}}{100~\text{GeV}}\right)^2 \frac{(1+r)^4}{r(1+r^2)}\left(1+0.07\log\frac{\sqrt{r}100~\text{GeV}}{m_{\widetilde l_R}}\right),
\label{relic-neutralino}
\end{eqnarray} 
where $M_1$ is the bino mass parameter. We note in passing that the limit $r\sim 1$ refers to the coannihilation regime. Eq.~(\ref{relic-neutralino}) cease to explain the relic density properly in such a scenario because of the presence of additional diagrams. It is also quite straightforward to see that the LEP limit on slepton masses, viz., $m_{\widetilde l_R}> 100$~GeV~\cite{LEP} and $r\lesssim 0.9$~leads to an overabundant universe. Therefore, a dominantly bino-like DM candidate must coannihilate with other MSSM particles. For pure Higgsino and wino DM, the relevant expressions for the relic density are~\cite{ArkaniHamed:2006mb}
\begin{eqnarray}
\Omega_{\widetilde H} h^2 &=& 0.1 \left(\frac{\mu}{1~\text{TeV}}\right)^2, \nonumber \\
\Omega_{\widetilde W} h^2 &=& 0.13 \left(\frac{M_2}{2.5~\text{TeV}}\right)^2.
\end{eqnarray}
In our study, to maximize the amount of coannihilation, we choose our spectrum as $m_{\widetilde q}>m_{\widetilde g}>m_{\widetilde t_1}>m_{\widetilde\chi_1^+}>m_{\widetilde\chi_1^0}$ where the mass difference is small amongst these fields.

It may be noted that the DM relic density is a serious concern in MUED because it allows less freedom in arranging the mass spectrum appropriately. However, as shown recently, a 5-dimensional UED can be embedded in a six-dimensional space-time with nested warping. The excitations of graviton in the sixth direction opens up new (co-)annihilation channels for the DM particle and as a result opens up new parameters spaces~\cite{Arun:2017zap}. In MSSM, the overabundance issue can be circumvented because of the freedom to choose soft mass parameters at the low scale. We note in passing that for both MUED and MSSM, the dark matter constraints are based on standard cosmology. For example, if the reheating temperature after inflation is lower than the freeze out temperature of  WIMP (LKP or LSP), then the relic abundance of the dark matter is reduced. This significantly relaxes~\cite{Roszkowski:2014lga} the stringent constraints coming from the overclosure of the universe.
\section{Benchmark points \label{benchmark}}
In this section, we carefully select suitable benchmark points (BP). Our intention is to generate a compressed mass spectrum which is, otherwise, difficult to search for at the LHC and to provide viable solutions in such situations. We also include constraints from direct dark matter searches as given by XENON1T~\cite{Aprile:2015uzo} experiment. For MUED benchmark, we consider $R^{-1} = 1.2$ TeV and $1.45$ TeV for BP1 and BP2, respectively. Note that, the latter choice of $R^{-1}$ results in a relic density within 3$\sigma$ of the experimentally observed value. Larger values of the same would give rise to an overabundant universe. Further, we choose the cut-off to be $\Lambda = 2R^{-1}$ unlike the conventional number $10R^{-1} - 20R^{-1}$. As mentioned earlier, we choose this value of $\Lambda$ to generate a sufficiently degenerate spectrum.

\begin{center}
\begin{table}[h]
\hskip 0.2 cm
\begin{tabular}{||c|c|c||c|c|c||} \hline\hline
Parameters  & BP1  & BP2  & Parameters & BP1  & BP2  \\
MSSM  & &  & MUED & & \\
\hline\hline
$M_1$ & 1.440 & 1.200 & $\Lambda R$ & 2 & 2  \\
$M_2$ & 1.380 & 1.200 & $R^{-1}$ & 1.2  & 1.45 \\
$M_3$ & 1.300 & 1.150 & $m_{G_1}$ & 1.285  & 1.553  \\
$A_t$ & 3.700 & 3.700 & $m_{D_1}$ & 1.254  & 1.515  \\
$\mu$ & 2.000 & 2.000 & $m_{U_1}$ & 1.254 &  1.515 \\
$\tan\beta$ & 20 & 20 & $m_{S_1}$ & 1.254  & 1.515  \\
$m_{\widetilde g}$ & 1.422  & 1.264 & $m_{C_1}$ & 1.254  & 1.515   \\
$m_{\widetilde q_{L}}$ & 1.470 & 1.310 & $m_{B_1}$ & 1.246 & 1.506 \\
$m_{\widetilde q_{R}}$ & 1.460 & 1.301 & $m_{T_1}$ & 1.244  & 1.499 \\
$m_{\widetilde t_1}$ & 1.409 & 1.225 & $m_{L_1}$ & 1.208 &  1.460 \\
$m_{\widetilde t_2}$ & 1.712 & 1.564 & $m_{Z_1}$ & 1.214 &  1.466 \\
$m_{\widetilde b_1}$ & 1.451  & 1.409 & $m_{W_1}$ & 1.214 & 1.466 \\
$m_{\widetilde b_2}$ & 1.597 & 1.457 & $m_{H_1}$ & 1.196  &  1.447 \\
$m_{\widetilde\ell_{L}}$ & 1.413 & 1.410 & $m_{d_1}$ & 1.247 & 1.507 \\
$m_{\widetilde\ell_{R}}$ & 1.406 & 1.405 & $m_{u_1}$  & 1.247 & 1.507 \\
$m_{\widetilde \tau_1}$ & 1.482 & 1.229 & $m_{s_1}$ & 1.247 &  1.507 \\
$m_{\widetilde \tau_2}$ & 1.532 & 1.285 & $m_{c_1}$ & 1.247 &  1.507 \\
$m_{\widetilde\nu_{L}}$ & 1.410 & 1.407 & $m_{b_1}$  & 1.247  & 1.507  \\
$m_{\widetilde\chi^0_2}$& 1.423 & 1.210 & $m_{t_1}$ & 1.258 &  1.516 \\
$m_{\widetilde\chi^{\pm}_1}$ & 1.388 & 1.210 & $m_{l_1}$ & 1.203 &  1.454 \\
$m_{\widetilde\chi^0_1}$ & 1.387 & 1.187 & $m_{\gamma_1}$ & 1.189 & 1.436 \\
\hline
\hline
$m_h$~(GeV) & 126.0 & 125.0 & $m_h$ (GeV)& 125.0 & 125.0 \\
$\Omega h^2$ & 0.109 & 0.110 &$\Omega h^2$  & 0.082 & 0.119  \\
$\sigma^p_{\text{SI}}$~(pb) & $8.01\times 10^{-10}$  & $1.21\times 10^{-10}$ & $\sigma^p_{\text{SI}}$~(pb) & $1.03\times 10^{-10}$ & $3.23\times 10^{-11}$   \\
$\Delta m_i $ (GeV) & 83.0 & 123.0 & $\Delta m_i$ (GeV)  & 96.0  & 117.0  \\
\hline\hline
\end{tabular}
\caption{The benchmark points corresponding to compressed spectra in the framework of minimal supersymmetric standard model and  minimal universal extra dimension are shown
in the left and right side of the table respectively. All the dimensionful input parameters and masses are expressed in units of TeV unless mentioned explicitly. Both the benchmarks
are subjected to the constraints from DM relic density and direct detection. Conventional notations are used for MSSM benchmark, whereas, notations for MUED  are explained in the text in sec.~\ref{mued}.}
\label{tab1:bp_comp}
\end{table}
\end{center}

For such choices, the typical MUED spectra follows a hierarchical structure such as $m_{G_1}>m_{Q_1}>m_{Z_1}>m_{\ell_1}>m_{\gamma_1}$. Instead of the doublet $Q_1$, the singlet KK quarks $q_1$ would also appear in the decay chain. Similarly, both $W_1$, {\it i.e.}, the KK $W$-boson and $Z_1$ would appear in the cascade. The relevant branching ratios are noted later in this section.

While choosing MSSM benchmark points, we keep the following hierarchy of masses: $m_{\widetilde q}>m_{\widetilde g}>m_{\widetilde t_1}>m_{\widetilde\chi_1^+}>m_{\widetilde\chi_1^0}$. The masses are sufficiently compressed so that LHC bounds do not work.  We must mention that our analysis remains valid even if, the hierarchy of masses  and the cascades are altered. The important point is to have sufficiently soft final state particles in the event. For example, choosing a lighter squark mass compared to the gluino mass would result in smaller cascade length. Hence, smaller number of soft tracks ($\xi$) would be observed in the final state. The values of $M_1$, $M_2$ and $\mu$ are mostly governed by amount of mixing in the lightest neutralino required for satisfying relic density and direct detection constraints. In case of BP1 in MSSM, both the lightest neutralino (LSP) and the lightest chargino (NLSP) are wino-dominated. Both the masses are governed by $M_2$ and hence, highly degenerate. The only accessible decay mode for the chargino is $\pi^+$ and the LSP. On the other hand in BP2, the LSP is bino dominated whereas the chargino continues to be wino dominated. Therefore, $\widetilde\chi_1^+$ mainly goes to $q\bar q^{\prime}\widetilde\chi_1^0$ via an offshell $W$. In addition, the overall mass splitting in the entire spectrum $\Delta m\equiv m_{\widetilde g}-m_{\widetilde\chi_1^0}$ has been kept somewhat different across the two MSSM benchmarks. This, as shown later, gives rise to slightly different $\xi$ distributions. A higher value of the stop trilinear parameter is required to ensure that the Higgs mass is around the experimentally allowed region. Therefore, $\widetilde t_2$ is much heavier than $\widetilde t_1$ due to the eigenvalue repulsion in the stop mass matrix and as a result $\widetilde t_2$ does not appear in the cascade. The input parameters, masses and numbers for comparison with various experimental data are listed in Table~\ref{tab1:bp_comp} for both the MUED and MSSM scenarios. The branching ratios of the various new physics particles for the two benchmark points are also shown in table \ref{tab2:branching}.

An important distinction between the MUED and MSSM benchmark points is the following. For our choice of parameters it turns out that level-1 KK $W$ or $Z$ boson will definitely appear in the cascade. This would give rise to soft leptons in the final state, which, in turn, gives rise to a smaller number of soft tracks as opposed to a purely hadronic cascade. In MSSM, if the sleptons are heavy, a larger number of soft tracks are expected because of the presence of hadronic modes~\footnote{For earlier studies regarding the discrimination of SUSY and UED at colliders see~\cite{Datta:2005zs,Battaglia:2005zf}}.

\begin{table}[h]
\centering
 \begin{tabular}{||c|c|c|c||} 
 \hline
 & Decay Channels  & BP1 & BP2 \\ [0.5ex] 
 \hline\hline
      & $G_1 \to Q_{i1} \bar{Q}_{i}$ + h.c. & 42.90\% & 42.91\%  \\ 
      & ~~~~~~~~$ q_{1i} \bar{q}_{i}$ + h.c. & 57.09\% & 57.09\% \\
      & $D_1 \to ~u W_1$ ~~~~~~~~~~ & 61.66\% & 61.68\% \\ 
 MUED  & $d Z_1$ & 32.38\% & 32.36\%\\
      & $Z_1 \to \nu_{1i} \bar{\nu}_{i}$ + h.c. ~~ & 50.34\% & 50.32\% \\
      & ~$\ell_{1} \bar{\ell}$ + h.c. & 49.66\% & 49.68\%\\
      & $W_1 \to \nu_{1\ell} \bar{\ell}$~~~~~~~~~~~~  & 49.66\% & 49.68\% \\
      & ~~$ \ell_1 \bar{\nu}_{\ell}$~~~~~ & 50.34\% & 50.32\% \\
      & $\ell_1 \to \gamma_{1} \ell$~~~~~~~~~~~~  & 49.66\% & 49.68\% \\
 \hline \hline
    & $\widetilde q_{L}\to \widetilde g~q_{L}$ & 60.0\% & 52.0\% \\
    & ~~~~~~~ $\widetilde{\chi_1^+}~q_L^\prime$               & 25.0\% & 32\% \\
    & $\widetilde q_R\to \widetilde g~q_R$ & 99.0\% & 91.0\% \\
    & $\widetilde g \to \widetilde t_1 \bar{c}$ + h.c. & 99.4\% & 99.7\% \\ 
    & $\widetilde t_1 \to \widetilde\chi^0_1 c $~~~~~~~~ & 2.96\% & 0.00478\%\\
 MSSM & $\widetilde\chi^{+}_1 b $~~ &  97.03\% & 93.83\% \\ 
    & $\widetilde\chi^{+}_1 \to \widetilde\chi^0_1 \pi^{+} $~~~~~ & 95.06\% & - \\
    & ~~$\widetilde\chi^0_1 q \bar{q'} $ & - & 65.76\% \\
    & ~~$\widetilde\chi^0_1 l^{+} \nu_l $ & 4.92\% & 34.24\% \\
\hline \hline
\end{tabular}
\caption{\label{tab2:branching} Some important branching ratios for MUED and MSSM benchmark points.}
\end{table}
%

\section{Collider Analysis {\label{analysis}}}
To probe a compressed spectrum, the typical search strategies rely on monojet/multijet signals associated with missing energy. As a result, one has to rely on an associated jet which recoils against the massive initial state system. Hence, the experimental event selection criterion is optimized with the requirement of having at least one jet with large transverse momentum ($p_T$) and a veto on isolated leptons and photons. Such a scenario, usually has a stable particle in the final state which gives rise to missing energy. Therefore, $\cancel{E_T}$ along with large $H_T$ where $H_T$ includes $p_T$ of soft tracks which are not inside any reconstructed objects
\begin{eqnarray}
H_T &=& \sum_{i=1..n~\text{jets}}p_T^i,
\end{eqnarray}
are useful variables. In addition, $M_{\text{eff}}$ is defined as
\begin{eqnarray}
M_{\text{eff}} &=& \sum_i p_T^i+\cancel{E_T}.
\end{eqnarray}

Missing energy content in a BSM scenario is typically larger than in SM. In case of a monojet $+ \cancel{E_T}$ search, the jet recoiling against the heavier non-standard particles would have much higher $p_T$ compared to the ones recoiling against $Z$ or $W$. As $H_T$ is defined as the scalar sum of the transverse momenta of all visible particles, it is also noticeably different compared to the usual SM backgrounds. Similarly, $M_{\text{eff}}$ is also a good discriminator in the same sense. However, all these variables are $p_T$ weighted and do not carry the full information of the event. A simple counting of the number of soft tracks gives an estimate of particle multiplicity in the event. For example, most of the BSM scenarios follow the typical hierarchical spectra where heavy colored particles are produced at the top of the decay chain through strong interactions and subsequently decay down to final state  stable objects which does not decay within the detector and as a result carry missing energy. Since the usual SM background is devoid of such a long cascade chain, the number of charged tracks is obviously much smaller as compared to the BSM signal. We reiterate that such tracks are associated with the primary vertex and are not part of any reconstructed objects. Associating the tracks with the primary vertex also gives robustness against pile-up and underlying events. We shall exploit this idea for different benchmark scenarios and compare the significance with and without the information of number of soft tracks. 
\begin{figure}[t!]
\includegraphics[height=5.0cm,width=7.0cm]{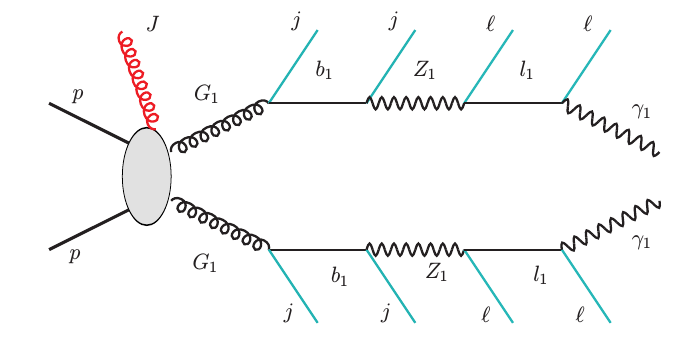}
\quad \includegraphics[height=5.0cm,width=7.0cm]{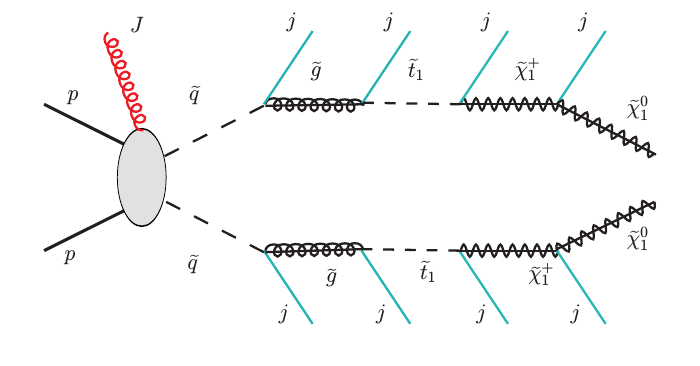}
\caption{A sample of the relevant topologies in both MUED and MSSM scenario. $J$ refers to the initial hard jet which triggers the event. $j$ and $\ell$ refers to soft jets and leptons which gives rise to tracks at the final state.}
\end{figure}

We have used \texttt{Feynrules-2.3}~\cite{Alloul:2013bka} to generate the MUED model file. For generating low scale MSSM spectrum, \texttt{SPheno-4.0.2}~\cite{Porod:2011nf} has been used. We have generated events using \texttt{Madgraph-2.3.3}~\cite{Alwall:2014hca} with the model files so generated at 13 TeV LHC\footnote{Event generation is also possible using {\tt CalcHEP} \cite{Pukhov:2004ca} for MUED ~\cite{Datta:2010us} and MSSM scenarios. The MUED model inside {\tt Pythia 8} has been recently implemented in \cite{Beuria:2017jez}.}. We have used {\tt NN23LO1} parton distribution function \cite{Ball:2012cx} which is available inside {\tt Madgraph} for all the processes. For MUED we have produced all the combination of KK gluons and quarks up to two jets. Similarly, all the relevant combinations of squarks (first two generations)-gluinos are also produced up to two jets. 

\begin{eqnarray}
p p &&\to G_1 G_1, G_1 Q_i(q_i), Q_i(q_i) Q_i(q_i), Q_i(q_i) \overline Q_i(\overline q_i) \nonumber \\
p p &&\to \widetilde q_i\widetilde q_j, \widetilde q_i\widetilde q_j^*, \widetilde g\widetilde g, \widetilde q_i\widetilde g, \widetilde q_i^* \widetilde g.
\end{eqnarray} 
A hard partonic cut $p_T^j>50$~GeV is used to produce the event files. For SM background, we have simulated $W$+jets, $Z$+jets, $t\bar t$+jets and QCD events. In principle, single top and SM di-boson production in association with jets would also contribute to the background. However, they are subdominant after implementing the event selection criteria. Both signal and background events are passed through \texttt{Pythia-8.2}~\cite{Sjostrand:2014zea} for hadronization and showering. In order to perform a semi-realistic detector simulation, we have used \texttt{Delphes-3.3}~\cite{deFavereau:2013fsa} with the default ATLAS card. Jets have been prepared using \texttt{Fastjet-3.2.1}~\cite{Cacciari:2011ma} with anti-$k_T$ jet algorithm \cite{Cacciari:2008gp} with jet radius of $R=0.5$. Tracks with $p_T>0.5$ GeV are selected with a $p_T-\eta$ dependent tagging efficiency. Tracks are also binned in $p_T$ as we will discuss later.
The relic density for all the BPs are calculated with the help of {\tt micrOMEGA-v4.1.4} \cite{Belanger:2014hqa}.
\subsection{Kinematic distributions}
\begin{itemize}
\item We prefer a harder cut (than the generation level cut) of $p_T^j>100$~GeV on a final jet. We also make sure that the final state is devoid of any isolated photons or leptons.
\begin{figure}[h]
\includegraphics[width=0.5\textwidth,height=6.0cm]{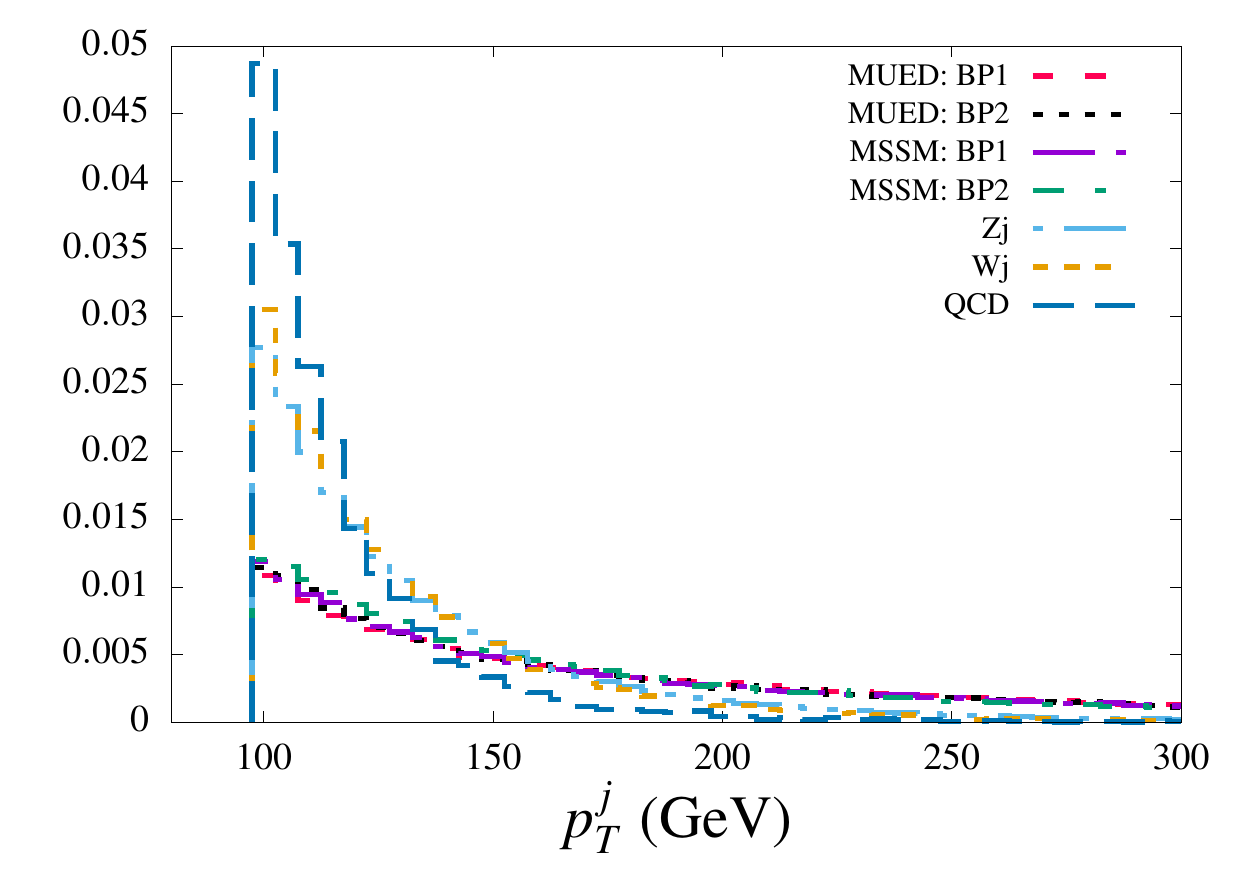}
\quad \includegraphics[width=0.5\textwidth,height=6.0cm]{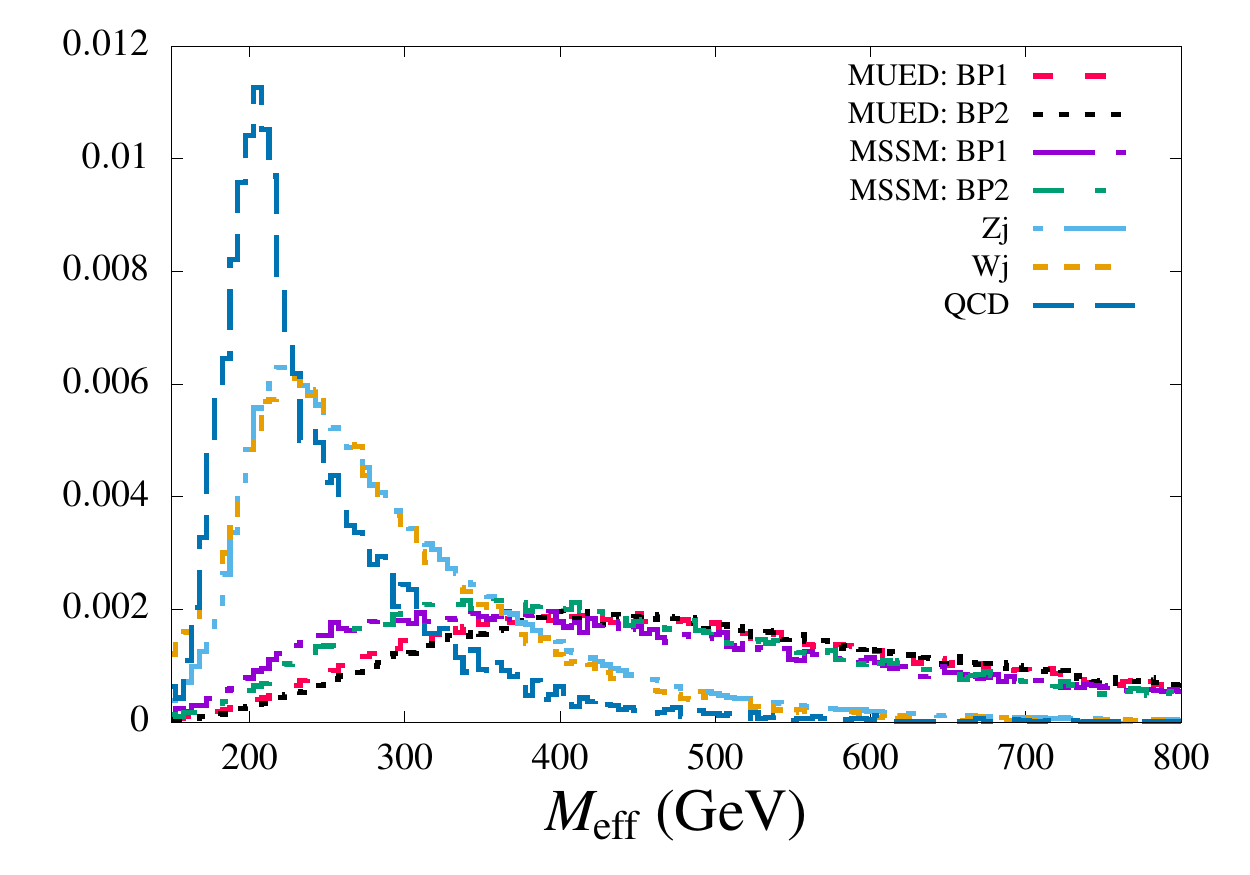}\\
\includegraphics[width=0.5\textwidth,height=6.0cm]{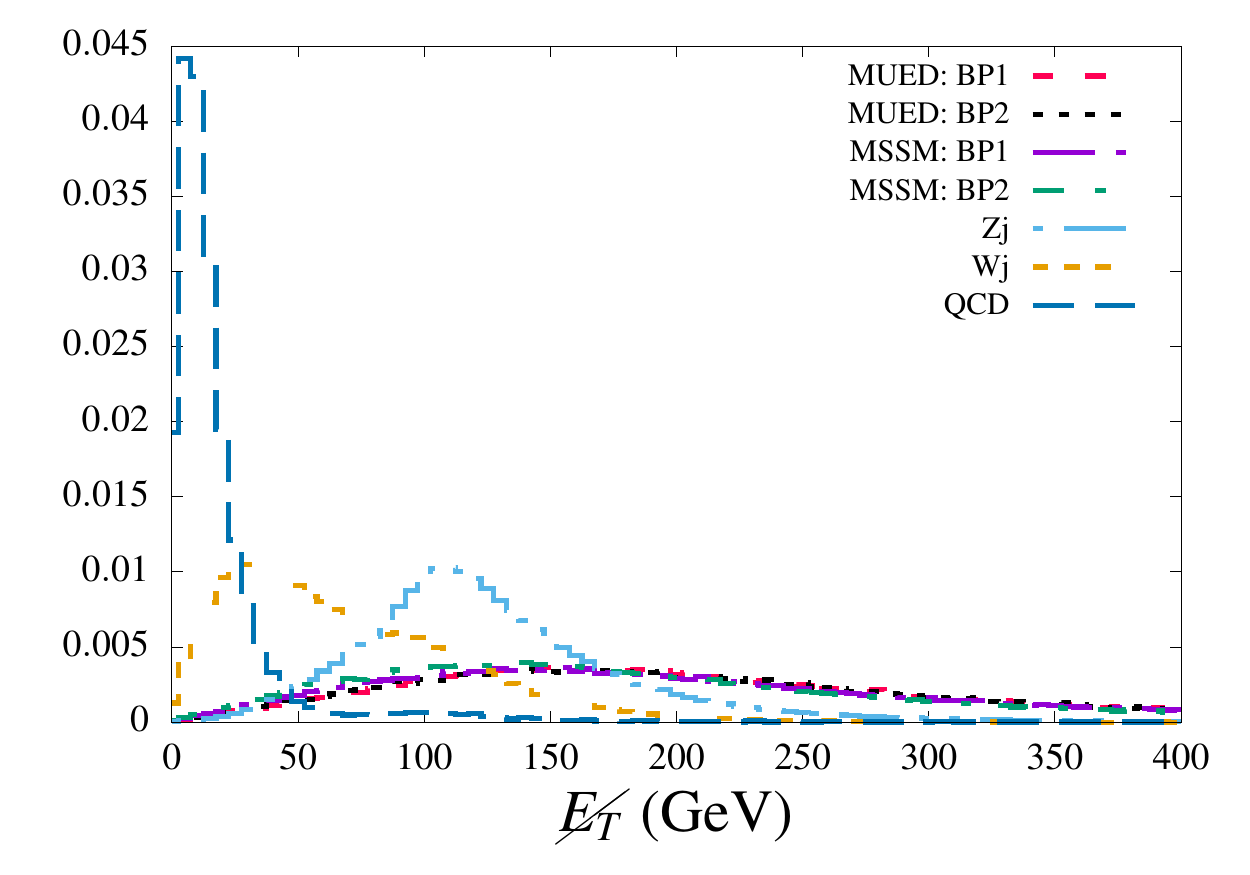}
\quad \includegraphics[width=0.5\textwidth,height=6.0cm]{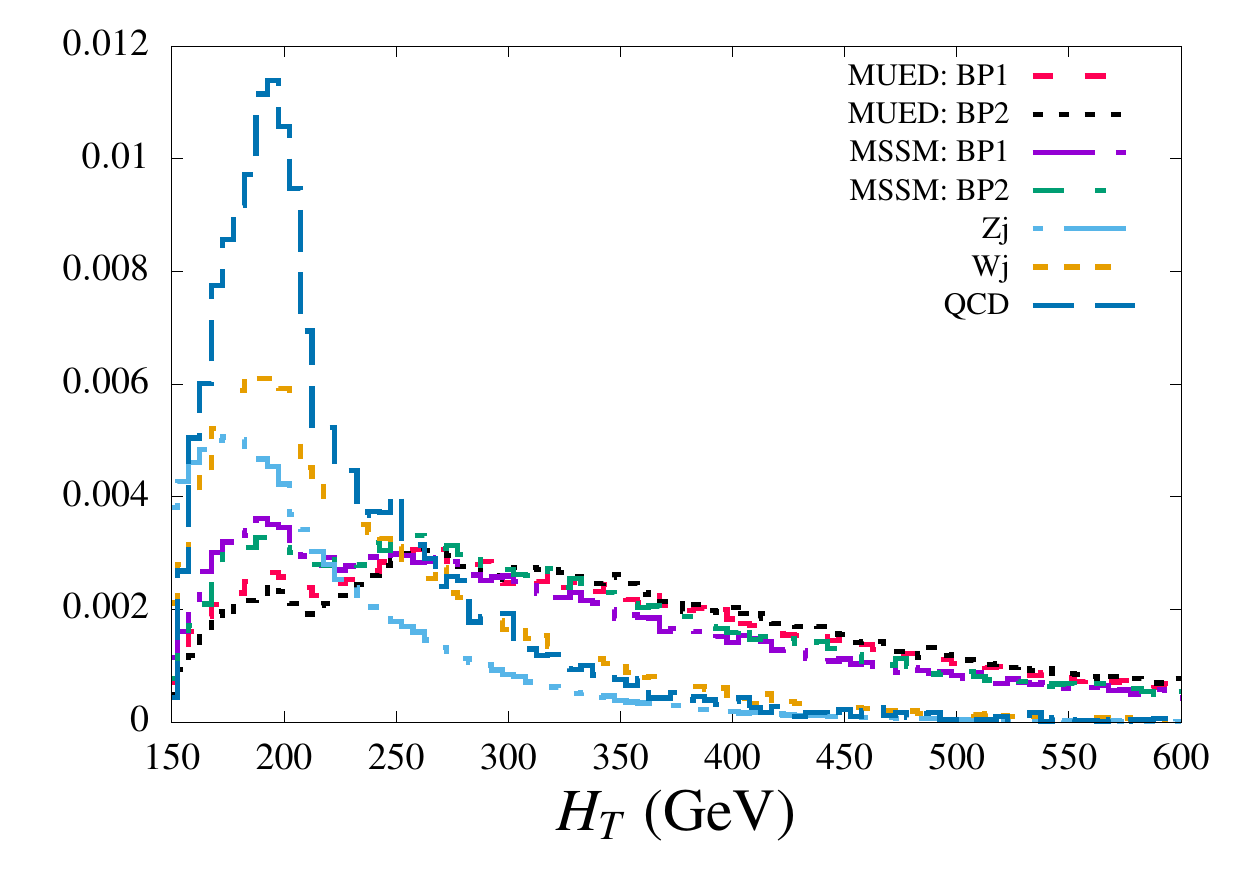}
\caption{\label{fig:usual_dis} Top row: Normalized distributions of the $p_T$ of the leading jet and $M_{\text{eff}}$ are plotted for the relevant SM backgrounds and four signal benchmark points corresponding to MUED and MSSM as shown in table~\ref{tab1:bp_comp}. 
Bottom row: Normalized distributions of $\cancel{E_T}$ and $H_T$ are shown for the same set of samples.}
\end{figure}
From fig.~(\ref{fig:usual_dis}), it is clear that $M_{\text{eff}}$ peaks at a higher value for the signal pertaining to our benchmark scenarios compared to the SM background. These distributions essentially help in selecting cuts required to dig out the signal usually buried under large background. For example, it is rather obvious from fig.~(\ref{fig:usual_dis}) that strong cuts, such as, $M_{\text{eff}}>800$~GeV and $H_T>400$ GeV could eradicate most of the SM background.

\item Moreover, the number of charged tracks in a cascade for both MUED and MSSM can be drastically different as compared to the SM processes. Longer cascades produce additional particles. But, the $p_T$ of those objects are less for a compressed scenario and as a result, such particles are too soft to give rise to any reconstructed objects. Compression to a certain extent ($\Delta m\sim 100$ GeV) and a longer cascade would naturally give rise to larger number of soft tracks. In a super-compressed scenario ($\Delta m\sim 25$ GeV), the tracks would be too soft and might not pass the threshold of track $p_T$ and as a result, efficiency would deteriorate~\cite{Chakraborty:2016qim}.  Of course, tracks are essential ingredients to fully reconstruct an object and have been used extensively for identification. However, our goal is to count the {\it number of soft tracks} which are associated with the primary vertex but, at the same time, are not used in any reconstruction process. This can be made sure by imposing the condition that angular separation between the monojet and the soft tracks are greater than the size of the jet itself. For simplicity, we will use `tracks' instead of soft tracks from now on.

\item An important point to note is that the number of tracks in an event is an infrared unsafe quantity \cite{Chakraborty:2016qim}. To elaborate, the properties of the tracks should not change when it passes through soft emission. Raw counting of the tracks as well as the parameters which depend on the track count are not infrared safe. However, charge particle count with a minimum $p_T$ is safer with respect to soft emissions~\cite{Gallicchio:2012ez}. Therefore, we have used $p_T$-binned tracks where the $p_T$ works as an effective cut on the number of tracks. As for robustness of this variable, we expect the $\xi$ distribution to vary by roughly 10\%~\cite{Gallicchio:2012ez} if different event generators such as Sherpa~\cite{Gleisberg:2008ta} and/or Herwig~\cite{Corcella:2000bw} are used. However, even then, significant discrimination between the signal and background can be observed in $\xi$.
Charged tracks are selected with $p_T^{\text{track}}>0.5$~GeV. Here, $\xi(0.5)$ implies the of number of the tracks inside the $p_T$ bin $0.5~\text{GeV} < p_T^{\text{track}} < 1.0~\text{GeV}$. Similarly, $\xi(1)$ denotes number of tracks between $1.0~\text{GeV} < p_T^{\text{track}} < 5.0~\text{GeV}$ and $\xi(5)\equiv p_T^{\text{track}}>5$~GeV.
\end{itemize}

\begin{figure}[h]
\includegraphics[width=0.5\textwidth,height=6.0cm]{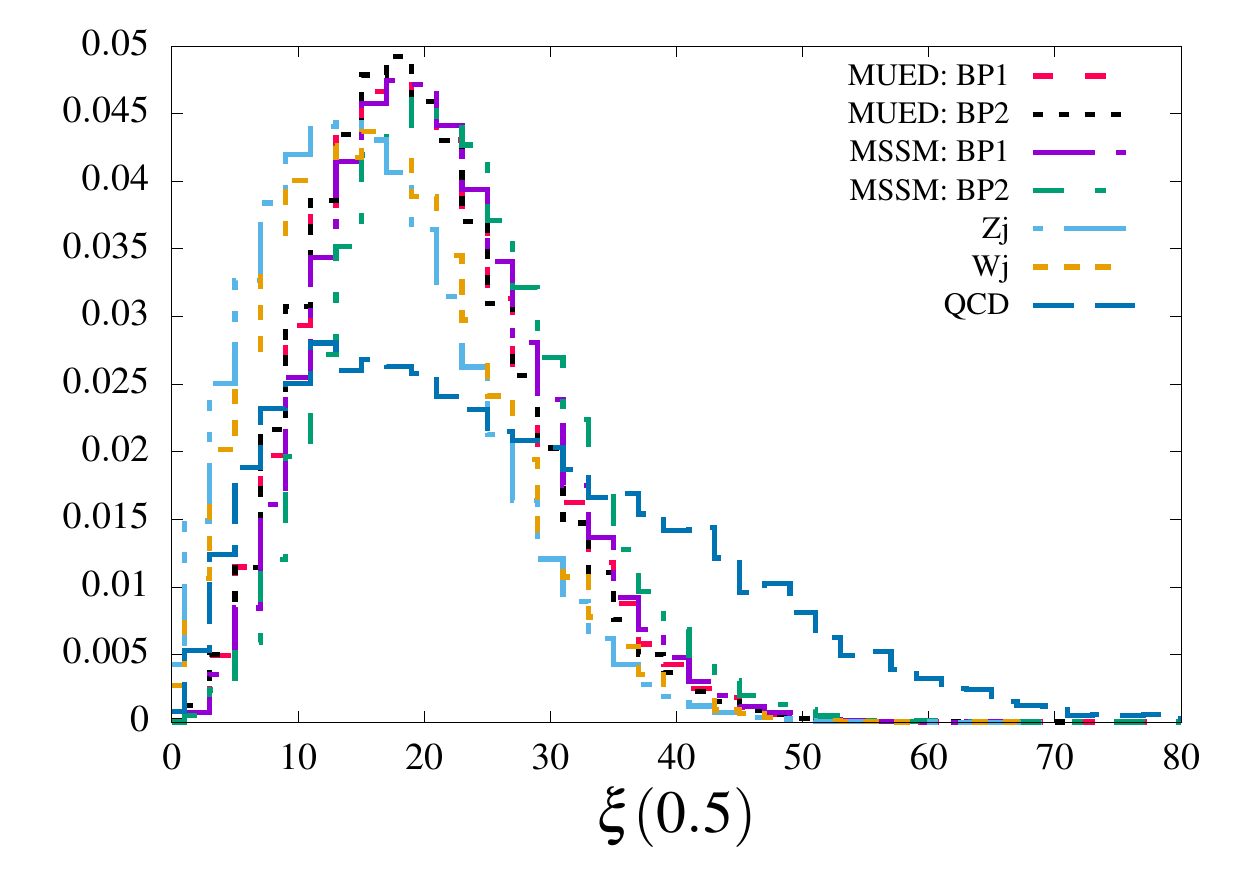}
\quad \includegraphics[width=0.5\textwidth,height=6.0cm]{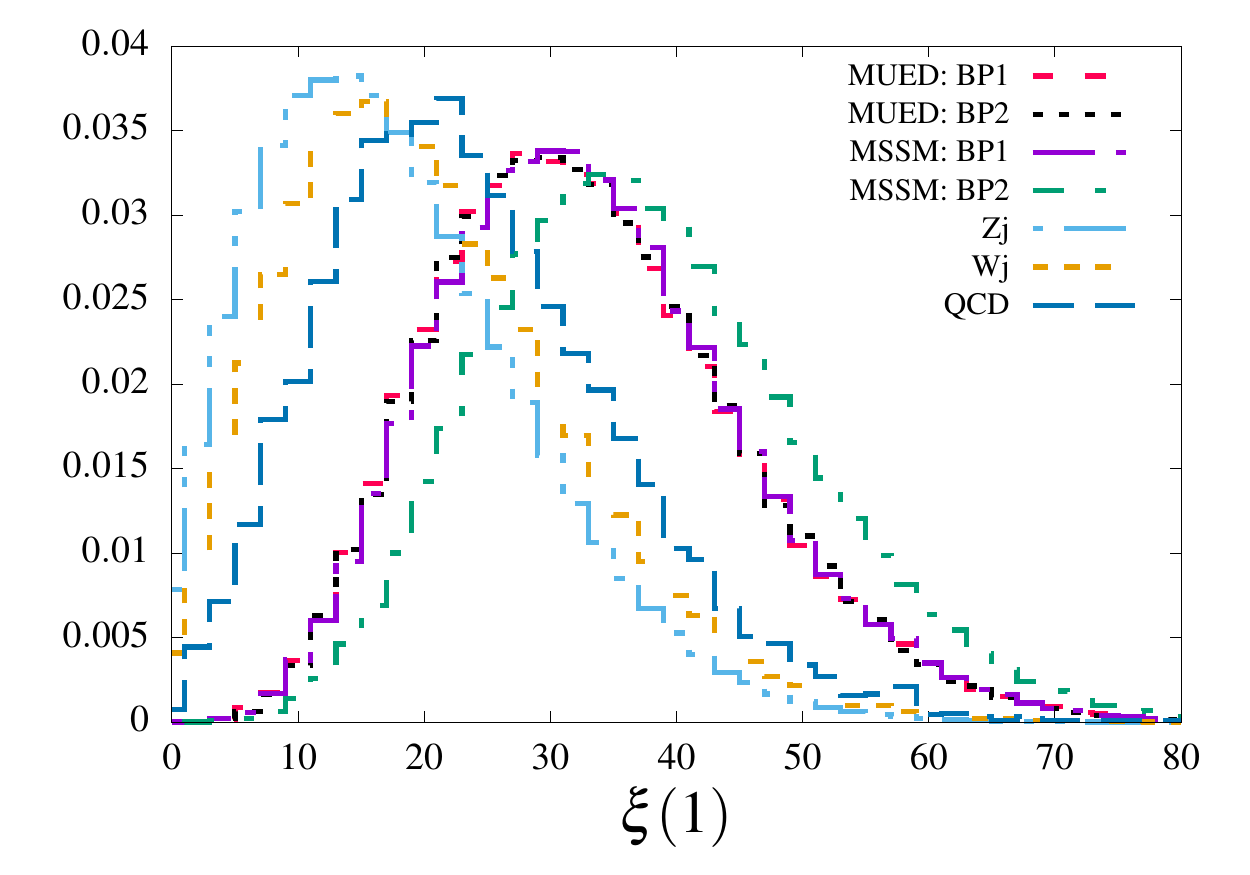}\\
\includegraphics[width=0.5\textwidth,height=6.0cm]{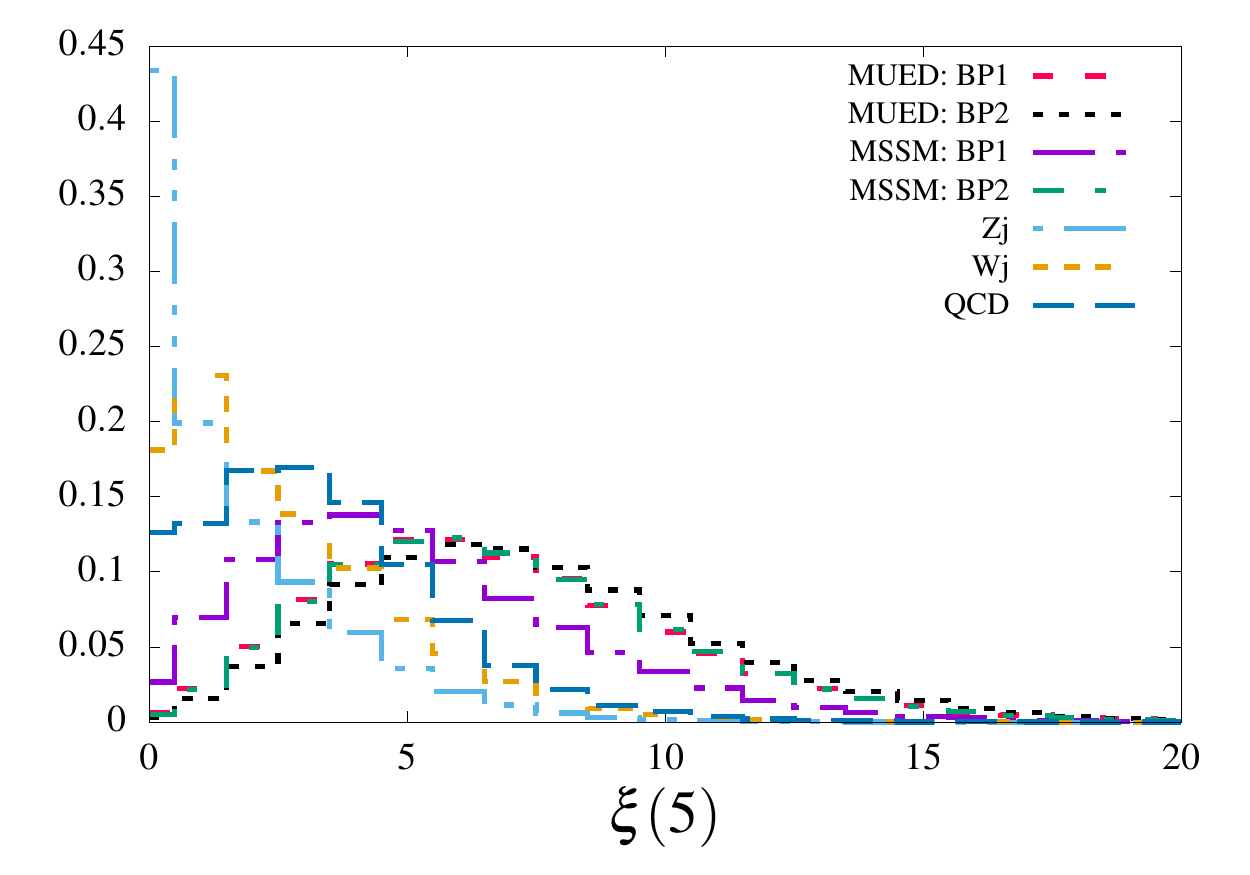}
\quad \includegraphics[width=0.5\textwidth,height=6.0cm]{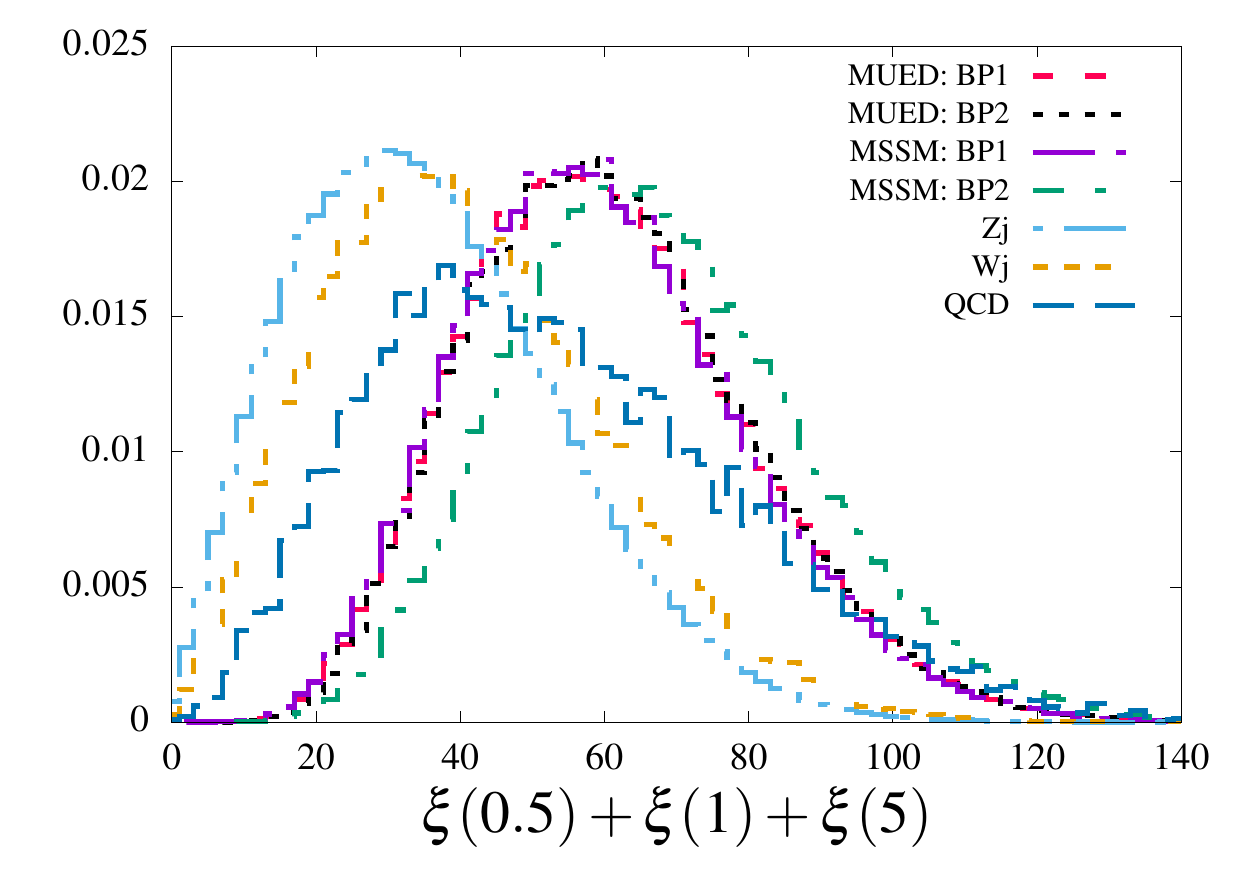}
\caption{\label{fig:track_dis} Distribution of number of soft tracks are displayed. The tracks are $p_T$ binned as described in the text. Plots are shown for all four signal benchmark points as well as for the SM backgrounds. The figures are normalized to unity.}
\end{figure}
The distributions of $\xi$ are displayed in fig.~(\ref{fig:track_dis}). QCD background gives a lot of soft tracks as can be seen from the figure at top-left. Other SM processes, in general, contain less number of tracks; or, in some cases, no track. Maximum number of tracks appearing in the new physics scenarios we consider here are in the $p_T$ range of $1-5$ GeV. Also, the figure of $\xi(1)$ shows that MSSM BP2 contains more number of soft tracks compared to BP1 as the degree of compression in BP2 is more. Background form strong interaction processes, i.e., QCD can be eliminated with a missing energy cut of $\cancel{E_T}>400$~GeV and $\Delta\phi(j,\cancel{E_T})>1.0$. The second cut ensures that $p_T$ of the jet is not aligned with MET, {\it i.e.}, missing energy is not sourced from jet-$p_T$ mismeasurement. Such a cut also reduces other SM backgrounds expect $Z$+jets. Hence, we will consider only $Z$+jets background for this work.

\begin{figure}[h]
\includegraphics[width=0.5\textwidth,height=6.0cm]{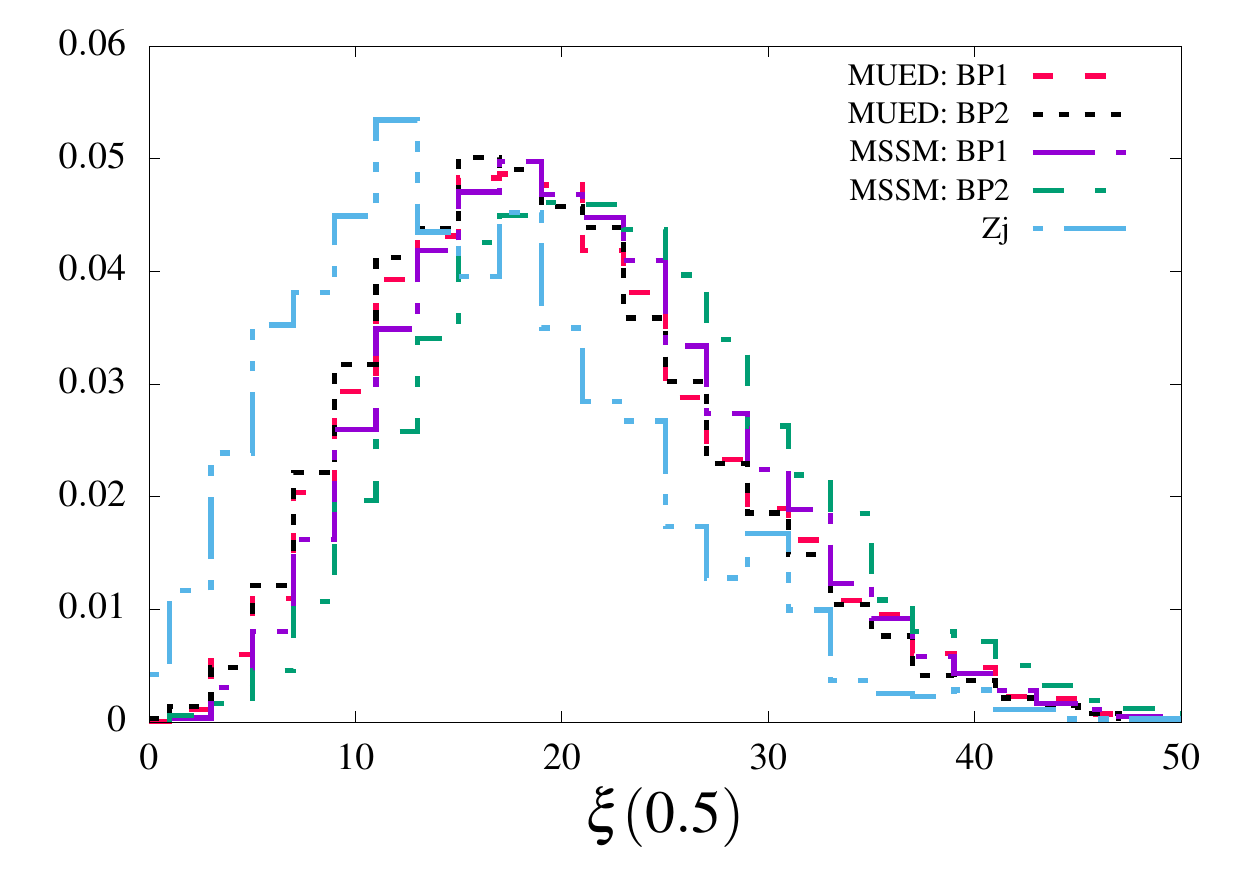}
\quad \includegraphics[width=0.5\textwidth,height=6.0cm]{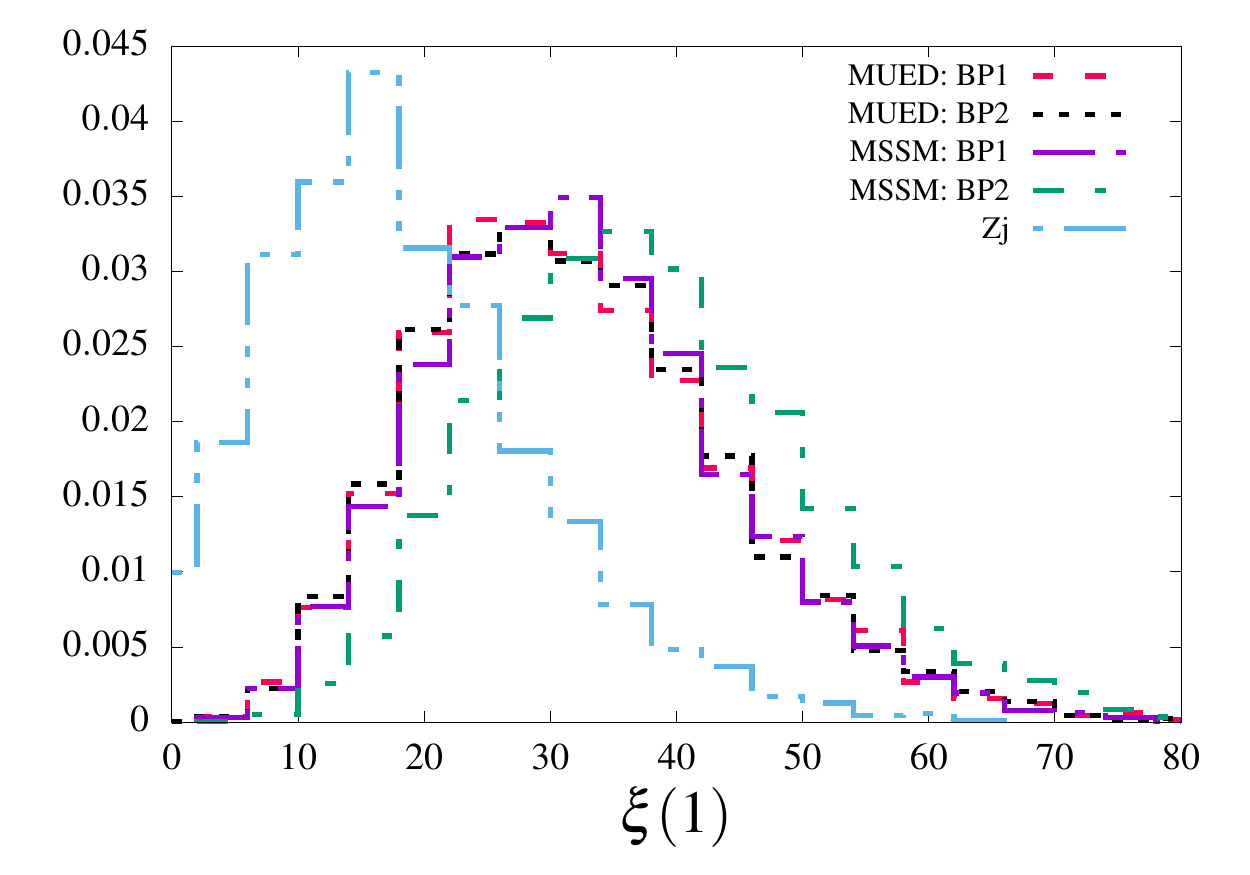}\\
\includegraphics[width=0.5\textwidth,height=6.0cm]{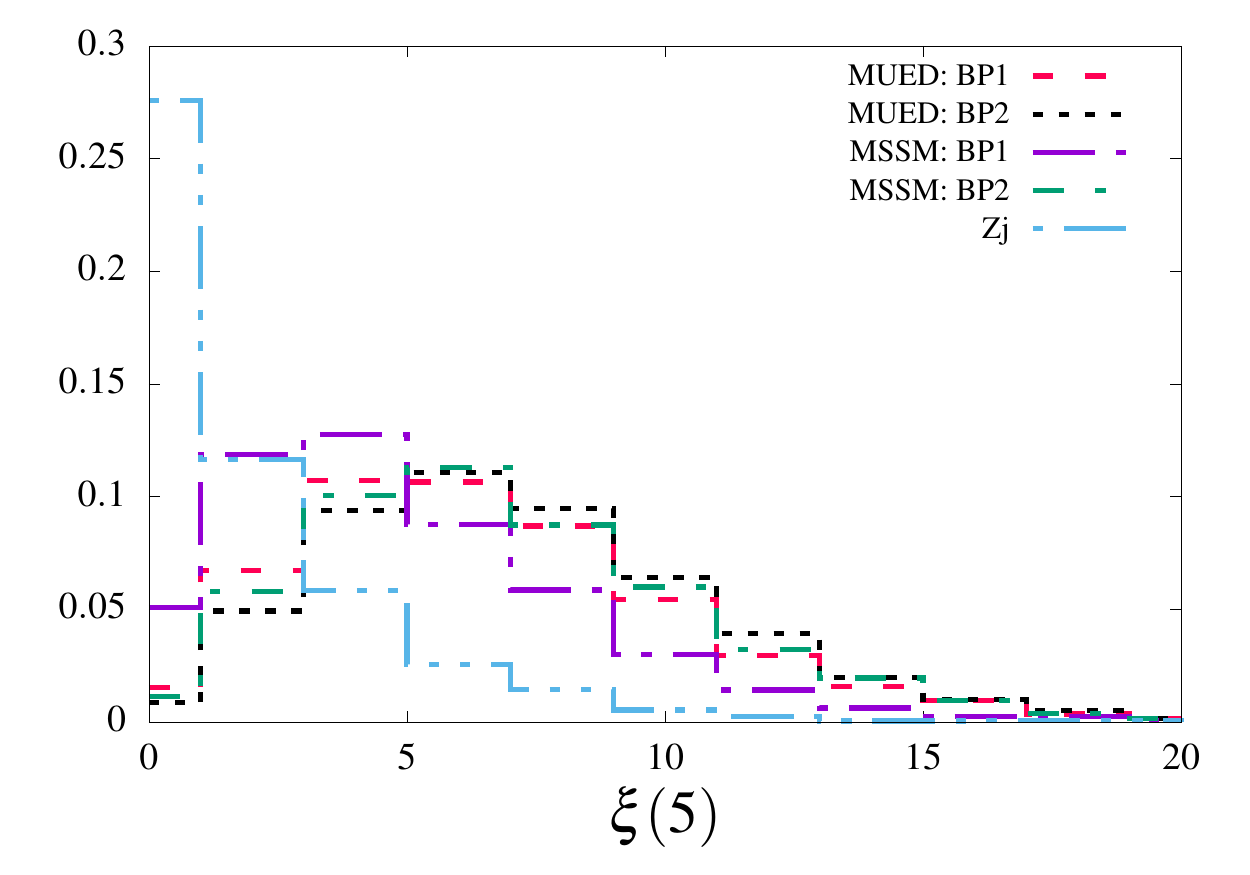}
\quad \includegraphics[width=0.5\textwidth,height=6.0cm]{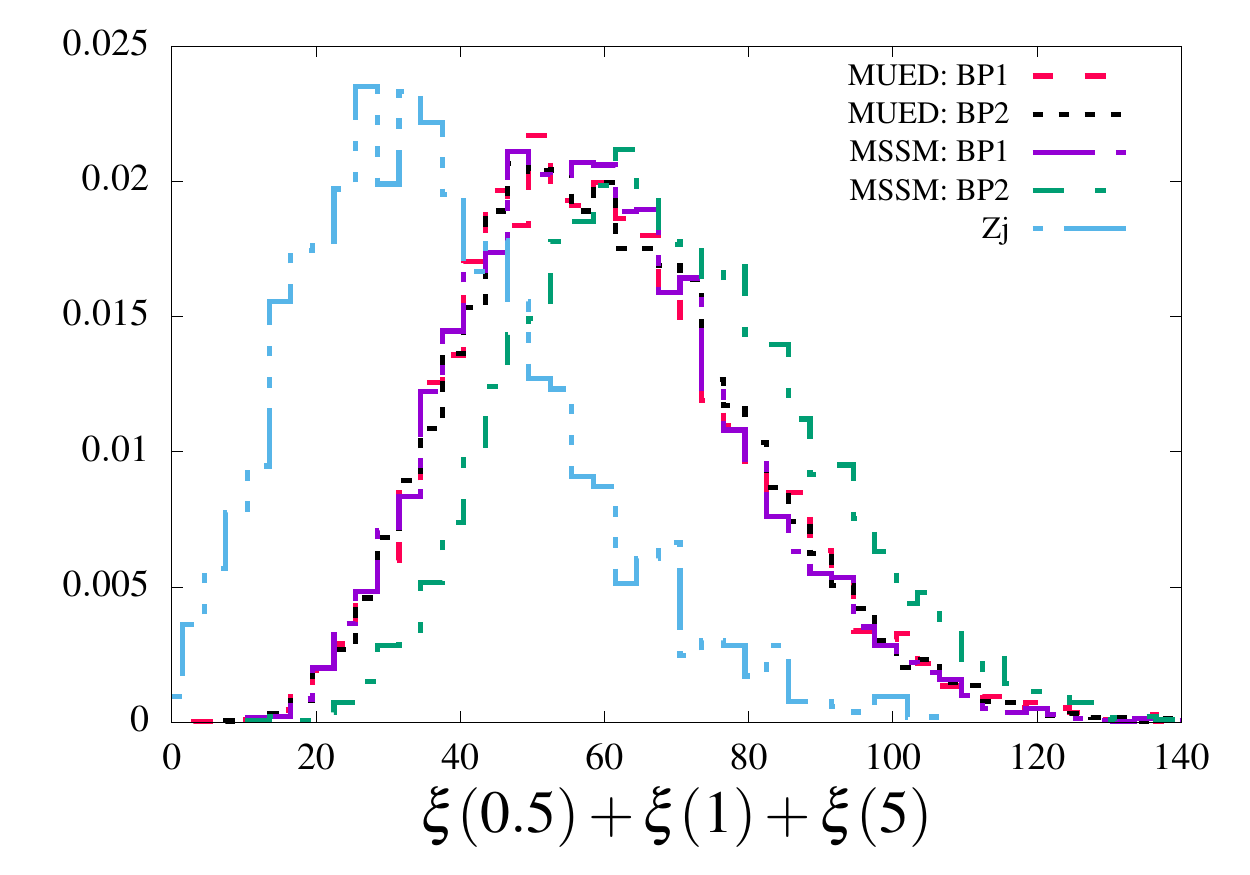}
\caption{\label{fig:trackm_dis} Number of soft track distributions after eliminating the SM QCD background by applying the cut {\bf C2}: {\it i.e.}, exclusive one jet with $p_T^j>100$~GeV, $\Delta\phi(j,\cancel{E_T})>1.0$ and $\cancel{E_T}>400$~GeV. A clear distinction can now be observed between the signal benchmarks and the SM background of $Z+$jets.}
\end{figure}
Below we note down the cuts used in the analysis:

\begin{itemize}
  \item {\bf{C0}}: A parton level cut of $p^{\text{patron}}_T > 50.0 $ GeV is used during the event generation in \texttt{Madgraph}.
  \item {\bf{C1}}: Events are selected with only single jet with $p^{j}_T > 100.0 $ GeV. Veto on isolated leptons and photons have also been applied.
  \item  {\bf{C2}}: $\cancel{E_T}>400$~GeV, $\Delta\phi(j,\cancel{E_T})>1.0$.
\end{itemize}

We have noted down the background and signal cross-sections after the use of all these cuts in table~\ref{tab:SM_bkg} and \ref{tab:SM_sig}. The modified distributions after implementing the cuts {\bf{C0}}, {\bf{C1}} and {\bf{C2}} are illustrated in fig.~(\ref{fig:trackm_dis}).

\begin{table}
\centering
 \begin{tabular}{||c|c|c|c||} 
 \hline
 SM Backgrounds & \bf{C0} & \bf{C1} & \bf{C2} \\ [0.5ex] 
 \hline\hline
 $Z+2~\text{jets}$ & $6.77\times 10^{5}$~fb & $1.1\times 10^{5}$~fb & 396.94~fb \\ 
 \hline
 $W+2~\text{jets}$ & $9.01\times 10^{5}$~fb & $5.68\times 10^{4}$~fb & 15.72~fb \\
 \hline
 $\text{QCD}$ & $10^8$~fb & $5.8\times 10^5$~fb & $\sim 0$ \\ [1ex]
 \hline
\end{tabular}
\caption{\label{tab:SM_bkg}Cross-sections for SM backgrounds after subsequent implementation of the cuts at 13 TeV.}
\end{table}
\begin{table}
\centering
 \begin{tabular}{||c|c|c|c||} 
 \hline
  Signals & \bf{C0} & \bf{C1} & \bf{C2} \\ [0.5ex] 
 \hline\hline
 UED: BP1 & 1423.3~fb & 511.1~fb  & 54.2~fb \\ 
 \hline
 UED: BP2 & 360.2~fb & 123.6~fb  & 13.18~fb \\
 \hline
 MSSM: BP1 & 92.3~fb & 36.94~fb & 4.48~fb \\ 
 \hline
 MSSM: BP2 & 207.9~fb & 75.76~fb & 7.0~fb \\ [1ex]
\hline
\end{tabular}
\caption{\label{tab:SM_sig}Cross-sections for the MUED and MSSM benchmarks at 13 TeV after subsequent implementation of the cuts as mentioned in the text.}
\end{table}
The reach of exploring new physics at the LHC increases with increasing centre of mass energy as well as integrated luminosity. However, at high luminosity the measurement of kinematic observables namely $p_T^j$, $\cancel{E_T}$ become extremely challenging due to the presence of large number of additional soft collisions occurring simultaneously with the hard collision, or pile-up. \texttt{Delphes} uses a fast jet area method to disentangle pile-up interactions with the interactions originating from high-$Q^2$ processes. This is quintessential for pile-up subtraction. To elaborate, \texttt{Delphes} identifies the primary vertex and removes all soft interactions which are outside a spatial distance $z$, set by the minimum resolution of the tracker. In a previous work~\cite{Chakraborty:2016qim} it was shown that the identification of primary vertex gives robustness against pile-up and the counting the number of soft tracks binned in $p_T$ is indeed a pile-up stable object. In this work we did not take pile-up into consideration.

\subsection{Cut based Analysis}

\begin{table*}[!h]
\renewcommand{\baselinestretch}{1}
\begin{center}
\begin{tabular*}{\textwidth}{@{\extracolsep{\fill}}lllcccccc||} \hline
     &   & $\cancel{E_T}$ & $H_T$ & $M_{\text{eff}}$ & $\xi(1)$ & $\xi(5)$ & Luminosity (fb$^{-1}$) \\ \hline \hline
& w/o track  & $400.0 $  & $700.0$  & $800.0$ & -- & -- & $127.23$\\ 
  BP1  &     &  & &  &  & &  \\ 
& with track  & $400.0 $  & $400.0$  & $800.0$ & $5.0$ & $4.0$ & $13.07$  \\ \hline \hline
& w/o track &  $400.0$ &  $700.0$ &  $800.0$ & -- & -- & $500.0$ \\ 
  BP2  &     &  & &  &  & &  \\ 
& with track  & $400.0$ & $500.0$  & $1000.0$ & $5.0$ & $5.0$ & $40.42$ \\\hline \hline
\end{tabular*}
\end{center}
 \caption{\label{tab:UEDsig} Luminosity required to achieve $5\sigma$ significance for the cut based analysis (with and without tracks) for MUED benchmarks.
 $\cancel{E_T}$, $H_T$, $M_{\text{eff}}$ are expressed in units of GeV.}
\end{table*}

In the usual cut based approach, we employ well-known $p_T$-based observables $\cancel{E_T}$, $H_T$, $M_{\text{eff}}$ to get appropriate signal significance over SM background. Furthermore, to see the effect of tracks as useful discriminating variable, we also provide significance numbers with and without tracks. A naive estimation from fig.~(\ref{fig:trackm_dis}) reveals that $\xi(1)$ and $\xi(5)$ are the most effective variables amongst tracks. Table \ref{tab:UEDsig} and \ref{tab:MSSMsig} show the luminosity required to achieve $5\sigma$ significance for the optimized set of cuts for two MUED and MSSM benchmark points respectively. It is clear that tracks turn out to be robust observables than other variables.

\begin{table*}[!ht]
\renewcommand{\baselinestretch}{1}
\begin{center}
\begin{tabular*}{\textwidth}{@{\extracolsep{\fill}}lllcccccc||} \hline
     &   & $\cancel{E_T}$ & $H_T$ & $M_{\text{eff}}$ & $\xi(1)$ & $\xi(5)$ & Luminosity (fb$^{-1}$) \\ \hline \hline
& w/o track  & $800.0 $  & $1000.0 $  & $700.0 $ & -- & -- & $162.86 $\\  
  BP1  &  &  & &  &  & &  \\ 
& with track  & $600.0 $  & $700.0 $  & $700.0 $ & $15.0 $ & $6.0 $ & $113.90 $  \\\hline \hline
&  w/o track &  $300.0 $ &  $700.0 $ &  $700.0 $ & -- & -- & $100.60 $ \\ 
  BP2  &     &  & &  &  & &  \\ 
& with track  & $300.0 $ & $700.0 $  & $700.0 $ & $15.0 $ & $6.0 $ & $57.34 $ \\\hline \hline
\end{tabular*}
\end{center}
 \caption{\label{tab:MSSMsig} Same as in table \ref{tab:UEDsig} for MSSM benchmark points.}
\end{table*}
We can see from table~\ref{tab:UEDsig} that 500 fb$^{-1}$ of integrated luminosity will be required to rule out BP2 in MUED. Similar outcome can hold true for the MSSM benchmarks as well.  A careful glance at tables~\ref{tab:UEDsig} and \ref{tab:MSSMsig} reveals that ruling out MSSM requires more luminosity at the LHC as compared to MUED. The cuts affect both the signal topologies in a similar manner. But, the MUED cross section (after cut) is much larger than the corresponding MSSM numbers as can be seen in table~\ref{tab:SM_sig}.
\subsection{Multivariate Analysis}
To optimize our search strategy and to show the relevance of adding soft tracks as a powerful discriminating variable, we perform a multivariate analysis using the Boosted Decision Tree (BDT) algorithm implemented in the Toolkit for Multivariate Analysis (TMVA)~\cite{Hocker:2007ht} within {\texttt{ROOT}}~\cite{Antcheva:2009zz} framework. A decision tree is essentially a classifier based on the decisions taken from a series of questions asked (or, conditions satisfied) in order to classify a set of data. In our case, this classification is whether a data is coming from signal or background. The questions are in the form of whether a data satisfies a particular cut or not and accordingly segregate the data. The tree starts from what is called a root node and finally, it arrives after a sequence of such segregations using some discriminating variables (which in our case, are the observables $M_{\text{eff}}$, $H_T$, $\xi$ etc.) applied to the data. Those variables are used that give the best separation between signal and background. However, results from a single tree are susceptible to statistical fluctuations. Therefore, it is better to take the majority vote from several trees which forms a forest. Boosting of a decision tree is also helpful to minimize such errors as it gives a larger weight to the missclasified events for the next iteration. It is also important to train the decision tree with a sample data as it refines the splitting criterion each time on repetition. This process is repeated until the best separation between the signal and background is obtained. Among many separation criteria, the Gini Index defined as $p \times (1 - p)$, where $p$ is the purity of the sample, is widely used.

We choose the following BDT parameters: number of trees in the random forest \texttt{NTrees}=400, maximum depth of the decision tree is chosen to be \texttt{MaxDepth=5} and the minimum percentage of training events in a leaf node is \texttt{MinNodeSize=2.5}\%. We keep all the other variables at their default values. In addition, we consider \texttt{AdaBoost} method for boosting the decision trees in the forest with the boost parameter \texttt{AdaBoostBeta}=0.5. 

Set A consists of the traditional variables such as $p_T^j$, $\cancel{E_T}$, $M_{\text{eff}}$ and $H_T$.  In addition, we have also taken the information of soft tracks in set B. 
\begin{table}[h!]
\centering
 \begin{tabular}{|c| c|} 
 \hline
 Set A & $p_T^j$,  $\cancel{E_T}$,  $M_{\text{eff}}$,  $H_T$ \\ [0.5ex] 
 \hline
 Set B & $p_T^j$,  $\cancel{E_T}$,  $M_{\text{eff}}$,  $H_T$,  $\xi(0.5)$,  $\xi(1)$, $\xi(5)$ \\
 \hline
\end{tabular}
\caption{\label{tab:MVA}To compare the change in signal significance without and with soft tracks we have introduced two sets of variables. Set A includes the conventional set of variables whereas Set B includes the $p_T$ binned track information along with the conventional variables.}
\end{table}

\begin{figure}
\centering
\includegraphics[width=15cm,height=7cm]{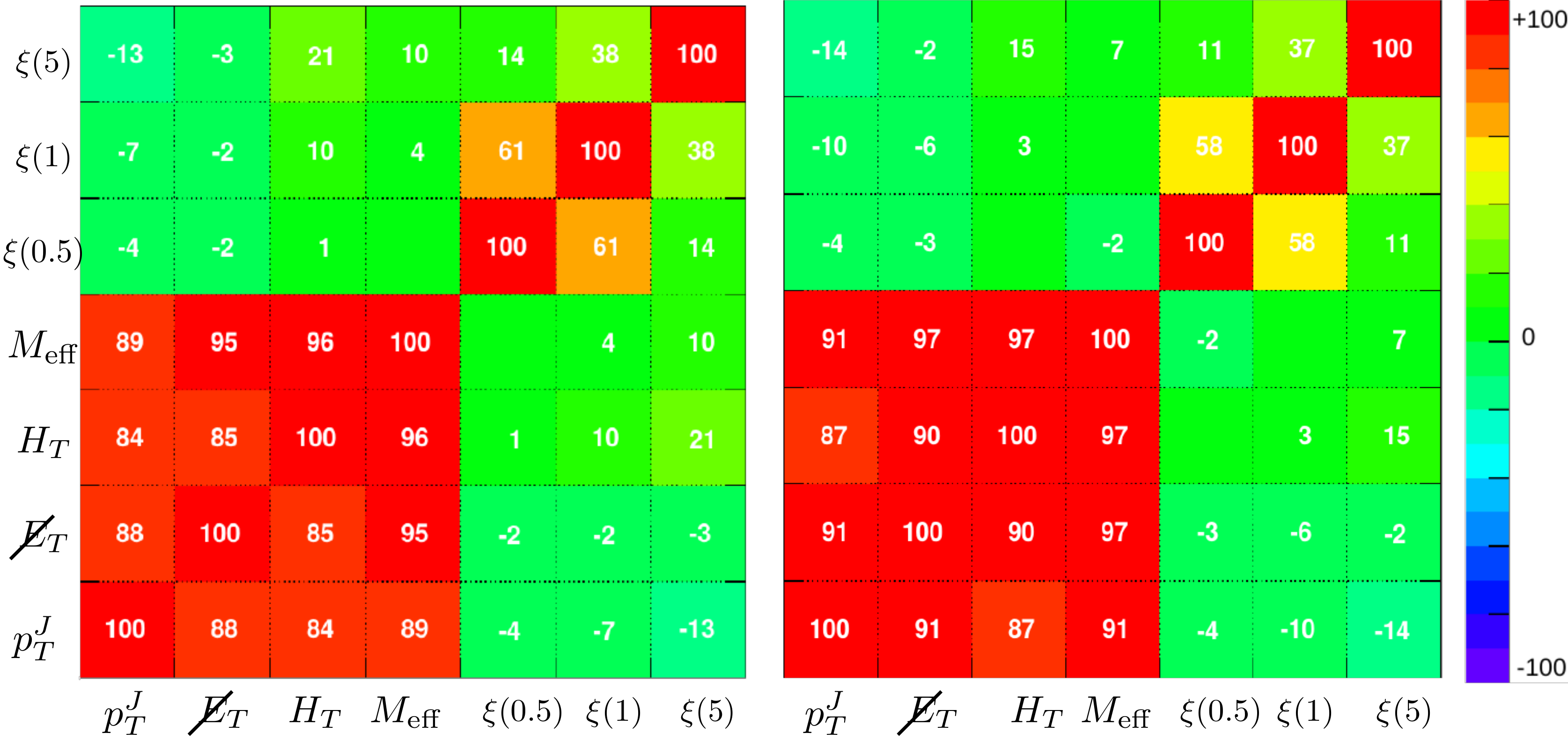}
\caption{\label{fig:corr_mva}Correlation between the variables for MSSM (left) and MUED (right). Clearly the $\xi$ variables are uncorrelated with the conventional variables which when taken into account should result in excellent $S/B$ as well as $S/\sqrt{S+B}$ improvements.}
\end{figure}
The correlations of our variables is shown in fig.~(\ref{fig:corr_mva}). The correlation between two such random variables, e.g., $X$ and $Y$ is measured with the correlation coefficient $\rho$ as~\cite{Hocker:2007ht}
\begin{eqnarray}
\rho(X,Y) &=& \frac{\text{cov}(X,Y)}{\sigma_X \sigma_Y},
\end{eqnarray}

where $\text{cov}(X,Y)=E(XY)-E(X)E(Y)$, $E$ is the expectation value. Fig.~(\ref{fig:corr_mva}) shows that the traditional variables are strongly correlated because of their very definition. However, the $\xi$ variables are uncorrelated with the rest and carry the information of particle multiplicity in an event. Therefore, in general it is expected to perform much better in a Multivariate Analysis. We note in passing that a slight correlation between $H_T$ and $\xi(5)$ can be observed because $H_T$ takes into account track $p_T$s.

\begin{figure}[h]
\includegraphics[width=0.5\textwidth,height=5.5cm]{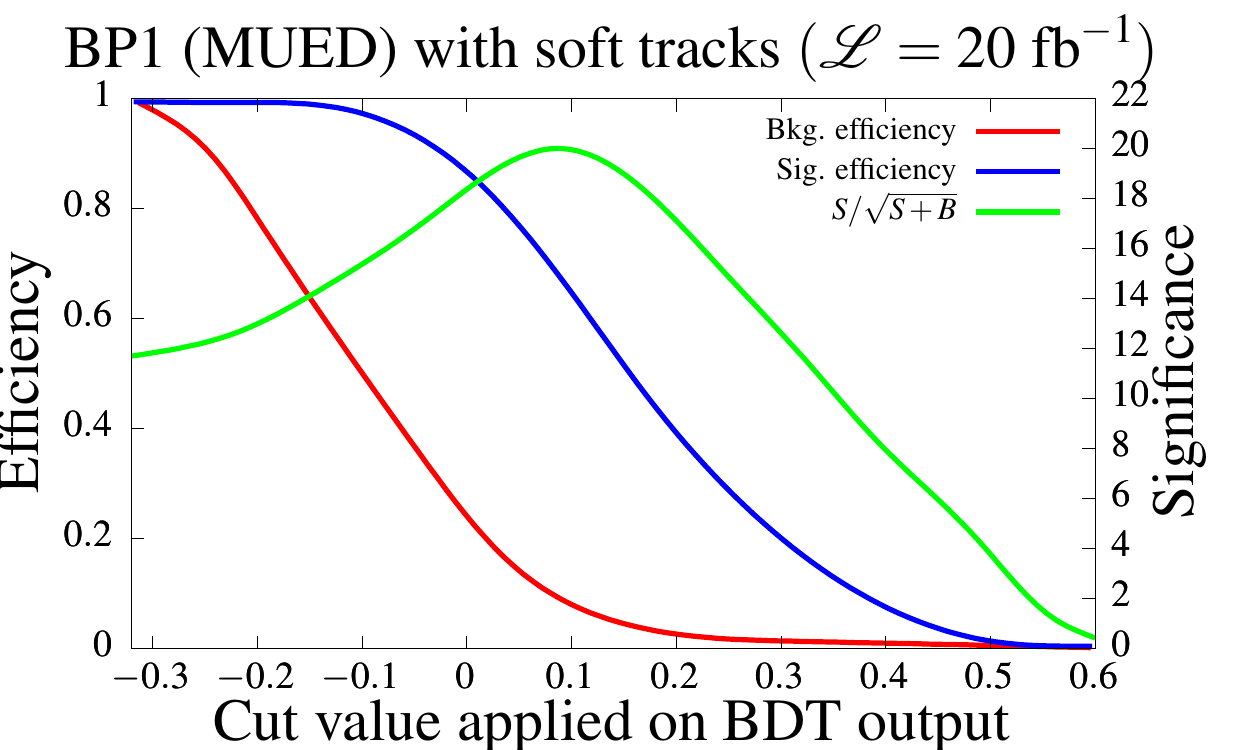}
\quad \includegraphics[width=0.5\textwidth,height=5.5cm]{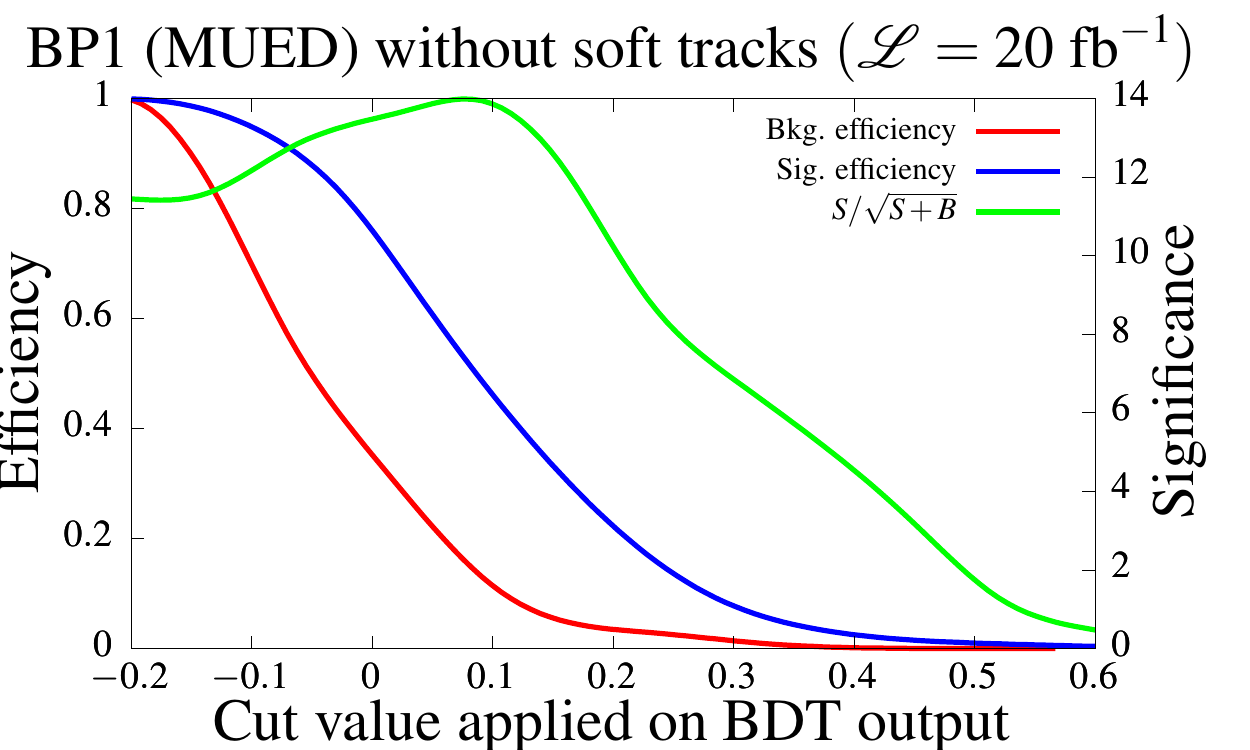}\\
\par
\includegraphics[width=0.5\textwidth,height=5.5cm]{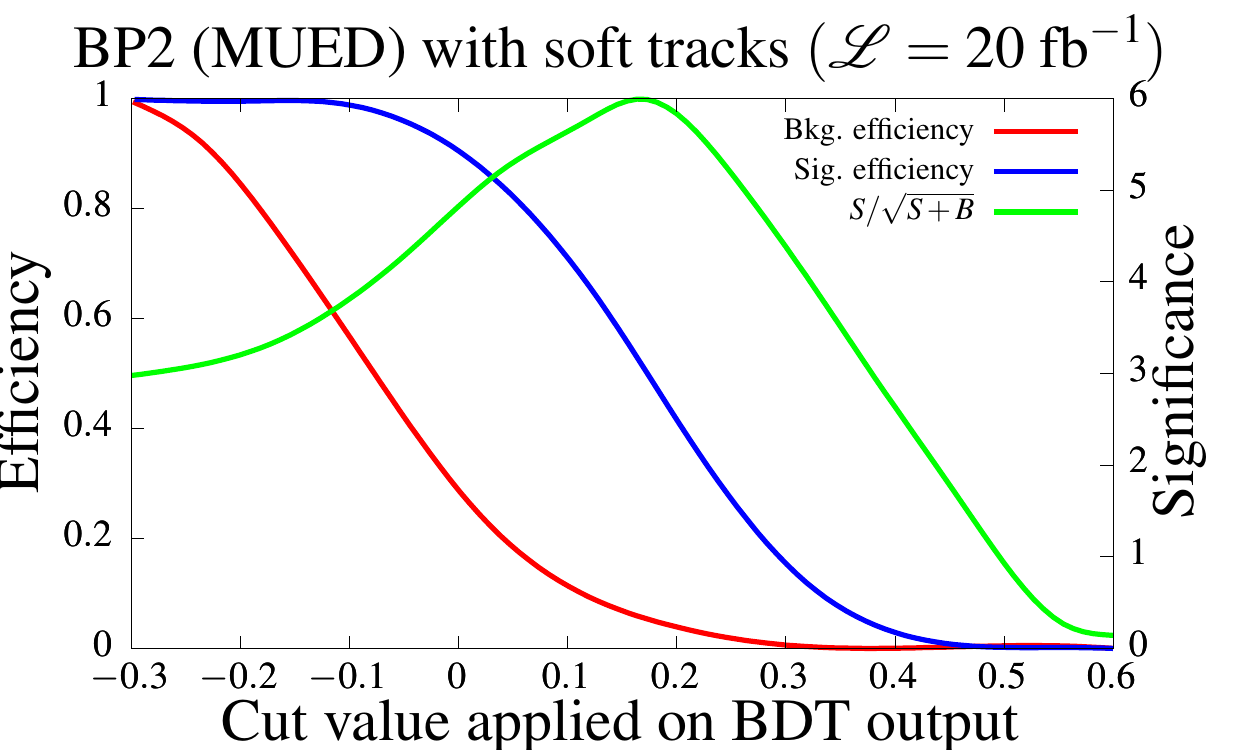}
\quad \includegraphics[width=0.5\textwidth,height=5.5cm]{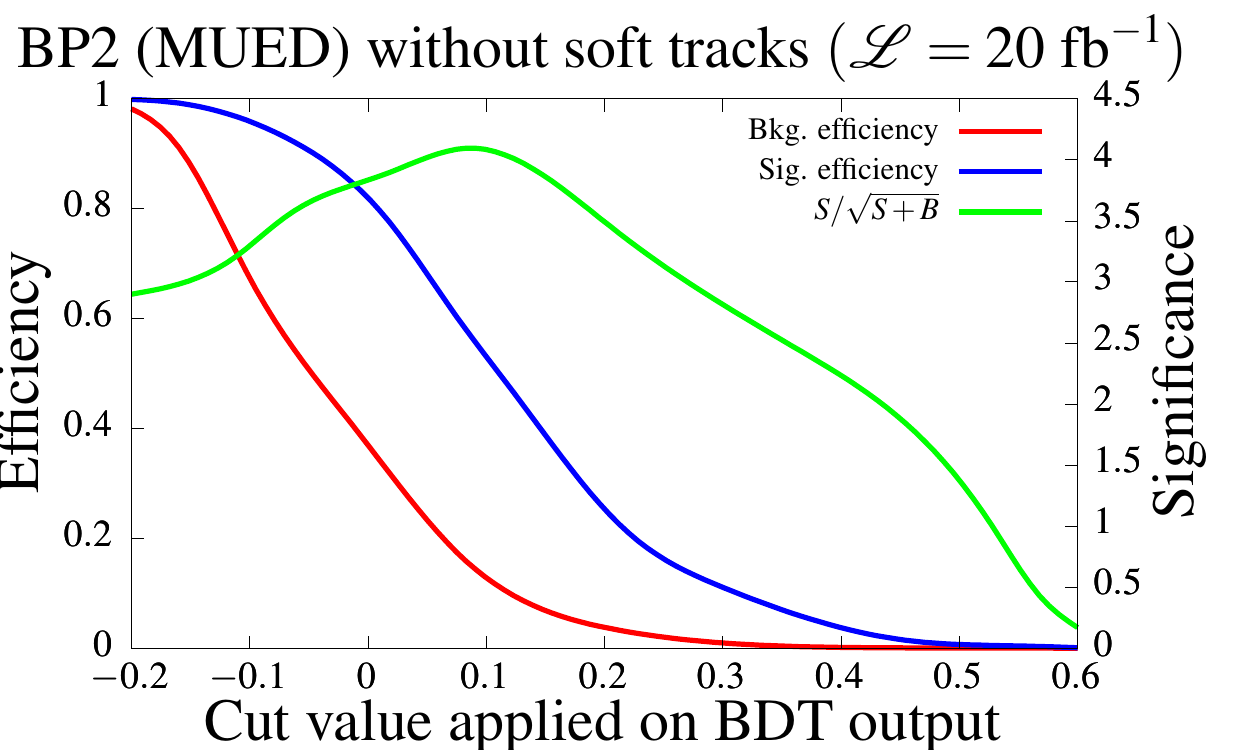}\\
\caption{\label{fig:ued_mva}We show the signal and background efficiencies with respect to the BDT output for MUED. The $y$-axis on the right hand side shows the signal significance pertaining to 20 fb$^{-1}$ of integrated luminosity as a function of the cut value on the BDT output.}
\end{figure}

In fig.~(\ref{fig:ued_mva}) we show the background and signal efficiencies along with the signal significance for two benchmark scenarios concerning MUED as a function of the BDT output. The BDT output depicts a mapping function of the $n$-dimensional phase space of the measured variables onto one-dimension. In general, one can consider any particular value of the BDT output as cut. However, one can see from fig.~\ref{fig:ued_mva} that a specific value of the BDT output gives highest significance. The left panel figures take into account the information regarding $p_T$ binned soft tracks whereas the right panel figures only include the conventional observables. It is clear that by taking $\xi$ into account, one gains in signal efficiency for the same background rejection. In other words, the purity of sample increases with the inclusion of $\xi$. Most importantly, the signal significance increases by roughly 50\% when $\xi$'s are taken into account for an optimized BDT cut value. To reiterate, BP2 marks the threshold for MUED since higher values of $R^{-1}$ are constrained from the over abundance of DM relic density. From fig.~(\ref{fig:ued_mva}) our analysis reveals that BP2 is all but ruled out with more than 5$\sigma$ significance with integrated luminosity as low as 20~fb$^{-1}$ (whereas, cut based analysis requires 40 fb$^{-1}$ of integrated luminosity for the same). 

However, in case of MSSM, the choice of soft mass parameters is not fixed unless we confine ourselves to a particular type of SUSY breaking. We took that liberty and tuned the soft parameters to obtain a rather compressed spectrum which is also compatible with cosmological observations. However, our goal is to show that a substantial increase in the signal significance can be obtained if soft tracks are used as an input in the MVA method.
\begin{figure}[h]
\includegraphics[width=0.5\textwidth,height=5.5cm]{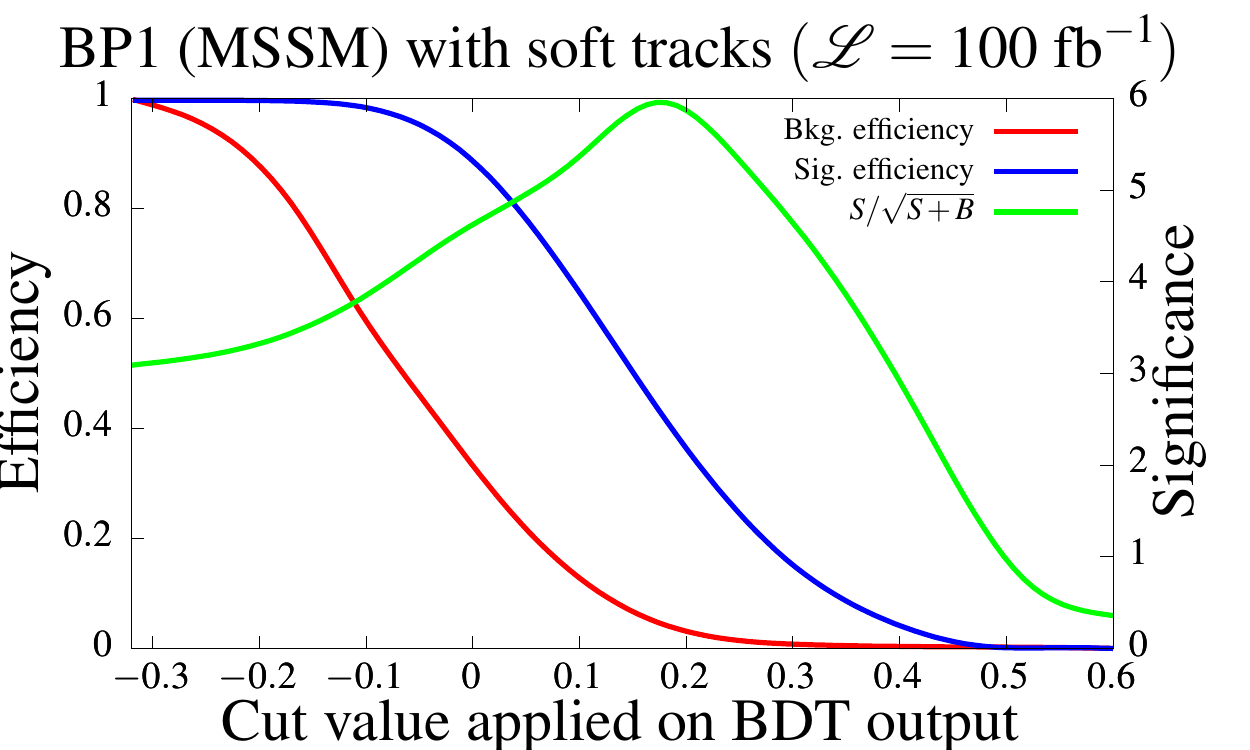}
\quad \includegraphics[width=0.5\textwidth,height=5.5cm]{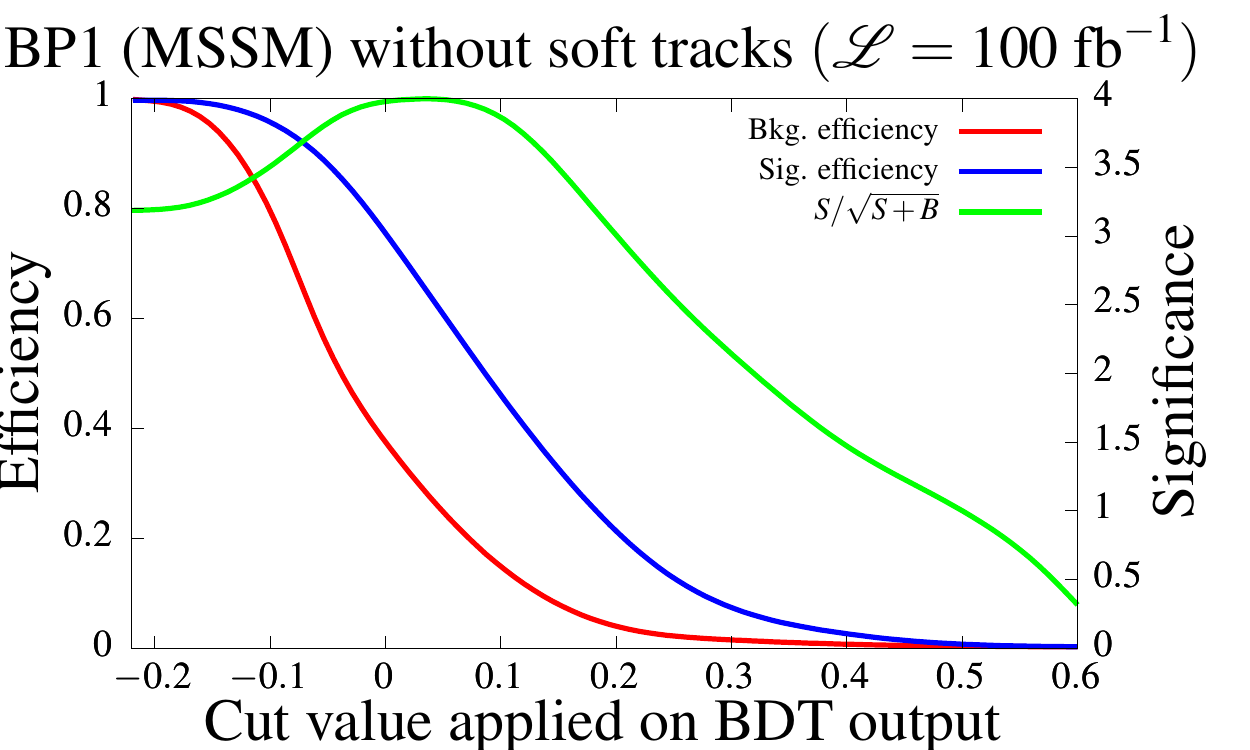}\\
\par
\includegraphics[width=0.5\textwidth,height=5.5cm]{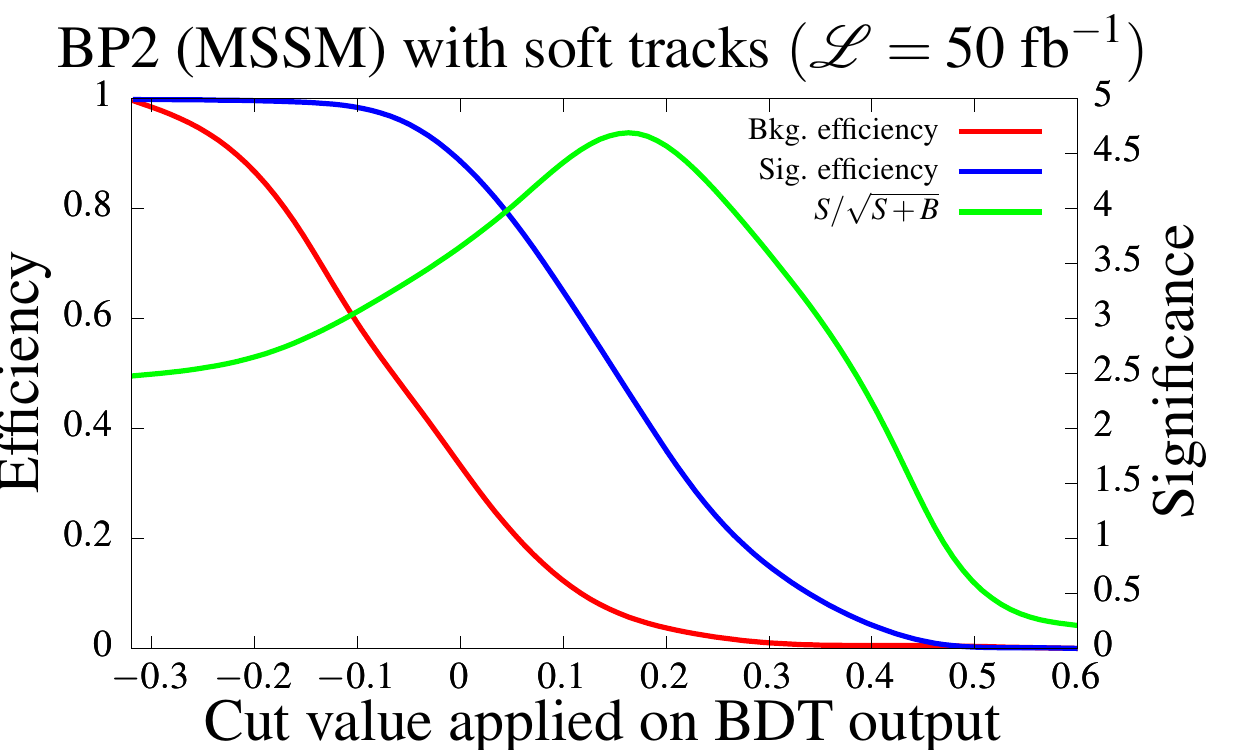}
\quad \includegraphics[width=0.5\textwidth,height=5.5cm]{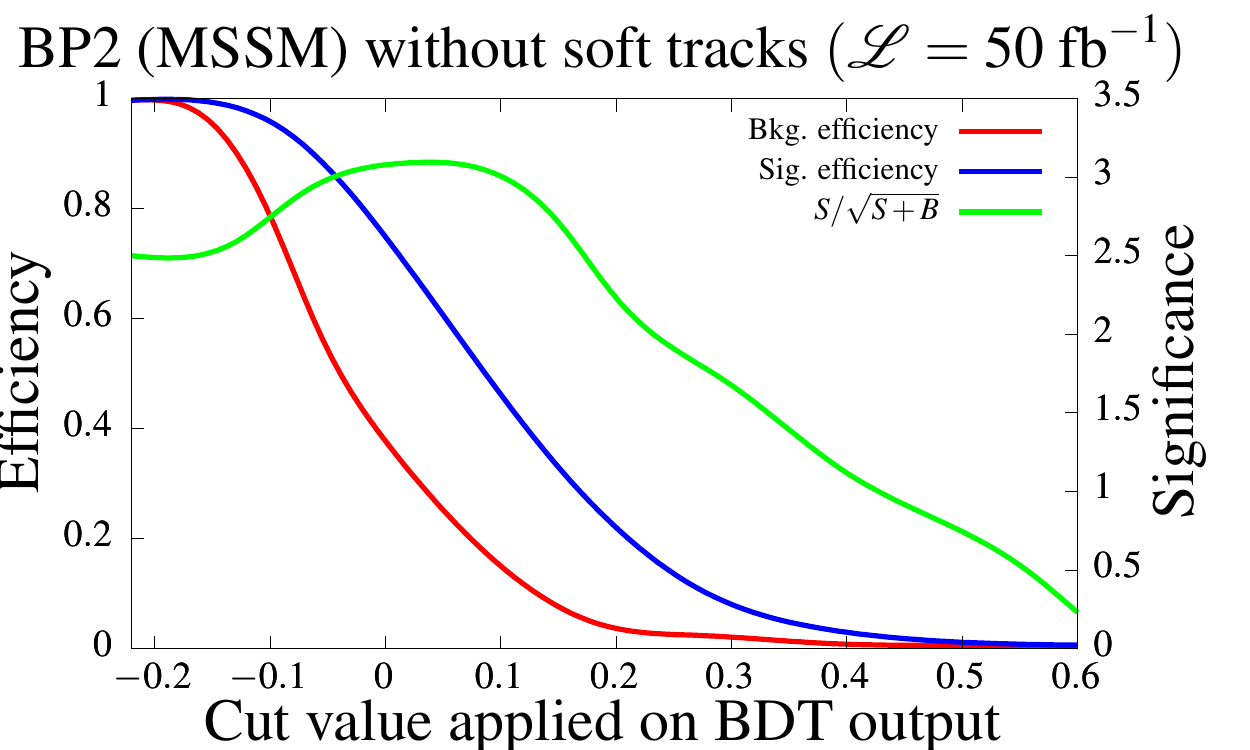}\\
\caption{\label{fig:mssm_mva} We show the signal and background efficiencies using the BDT output for MSSM. The $y$-axis on the right hand side shows the significance as a function of the cut value on the BDT output for a given luminosity. The figures in the top (bottom) row are with 100 fb$^{-1}$  (50 fb$^{-1}$) integrated luminosity.}
\end{figure}
Our results has been elucidated in fig.~(\ref{fig:mssm_mva}). It is obvious that signal significance increases with Set B cuts. BP1 for MSSM can be probed with conventional observables at a level of 4$\sigma$ significance with the integrated luminosity of 100~fb$^{-1}$ data. Remarkably, the signal significance can be enhanced to roughly 6$\sigma$ with $\xi(0.5),~\xi(1),~\xi(5)$ for the same integrated luminosity. This points towards an 50\% increase of the signal significance.
\section{Conclusion \label{conclu}}
\begin{itemize}
 \item In situations where the mass spectrum of a new physics scenario is almost degenerate, namely compressed spectrum, provides a challenge to extract the signal from the overwhelmingly large SM background. The main irreducible background in such case is $Z$+jets. Such a huge background reduces efficiency and hence, results in weaker bounds on exotic particle masses. In this work, we have considered two benchmark points in the MUED and MSSM frameworks with compressed spectrum. Our choice of benchmark points are consistent with all the present collider and DM bounds. Such scenarios are conventionally studied with an associated hard jet that helps the rest of the system (hence, final decay products) gain enough recoil energy to register in the detector. In our case, the final state objects could not be reconstructed because the compression imparts very low $p_T$. The goal of this paper is to show that by counting (soft) particle multiplicities (as tracks) along with a hard jet in the final state increases the signal sensitivity and enhances the significance. 

 \item The observables $p_T$, $H_T$, $M_{\text{eff}}$ are all energy weighted and therefore, highly correlated. We particularly emphasize that $p_T$-binned track observables ($\xi$) are uncorrelated and carries more independent information about an event. Therefore, inclusion of $\xi$ from the soft objects could provide a very effective handle for examining quasi-degenerate masses. The two representative benchmark points for MUED and MSSM scenarios are exposed to both cut-based and multivariate analysis. We observe significant improvement in the cut-based analysis when the $\xi$ variables are added with the traditional variables. To illustrate, BP2 in MUED can be ruled out by $5~\sigma$ significance with 1/10th of integrated luminosity when track variables are taken into account. We then perform a state-of-the-art multivariate analysis to optimize our search strategy. Our conclusion is that MVA performs much better as it could rule out more parameter space effectively with moderate luminosity.
 
 \item Constraints from DM relic density and collider searches has left little room for MUED. However, LHC sensitivity is rather weak in compressed scenarios. We show in our analysis that such a compressed version of MUED with $\Lambda R=2$, can be easily ruled out with $\sim 20~\text{fb}^{-1}$ of data which is already collected at the LHC at 13 TeV center-of-mass energy. We show this with our choice of BP2 for MUED with $R^{-1}=1.45$~TeV. Higher values of $R^{-1}$ leads to an overabundant universe under the assumption of standard cosmology.

\item Finally and perhaps most importantly, there is a straightforward way to estimate the number of soft tracks binned in $p_T$ in a data-driven way. We have checked that using $Z\to \mu^+\mu^-$ associated with a jet events~\cite{Chakraborty:2016qim}, one has to count the number of soft tracks which are not inside the jet as well as not inside the muons. This gives an excellent estimate for all the $\xi$ variables. 
\end{itemize}

Therefore, we advocate that information on number of soft tracks can be used as a powerful variable to distinguish compressed scenarios appearing in various frameworks. Particularly, performance of MVA technique becomes much better if such track variables are taken into account.
\section{Acknowledgment}
We thank Debajyoti Choudhury and Debjyoti Bardhan for helpful discussions. SC would like to thank Tuhin S. Roy and Amit Chakraborty for discussions. SC also acknowledges the hospitality of Korea Institute for Advanced Study during which the final part of the project was completed. SN acknowledges Dr. D. S. Kothari Post Doctoral Fellowship awarded by University Grant Commission (award letter no. PH/15-16/0073) for financial support.

\end{document}